\documentclass[12pt]{article}
\usepackage{amsfonts}
\usepackage{amsmath, amsthm}
\usepackage{epsfig}
\usepackage{algorithm}
\usepackage{algorithmic}
\usepackage{epic}
\usepackage[a4paper]{geometry}
\usepackage{enumitem}
\usepackage{dsfont}
\usepackage{rotating}
\usepackage{lscape}

\usepackage{amsmath,verbatim,color,amssymb,epsfig}
\usepackage{bm}
\usepackage{amsfonts}
\usepackage{epsfig}
\usepackage{changepage}
\usepackage{multirow}
\usepackage{graphicx}
\usepackage{array}
\usepackage[table]{xcolor}
\definecolor{Gray}{gray}{0.80}
\usepackage{lscape}
\usepackage{rotating}
\usepackage[round,semicolon,authoryear]{natbib}
\usepackage[normalem]{ulem}
\usepackage[colorlinks=true,urlcolor=blue,citecolor=purple,linkcolor=blue,bookmarks=true]{hyperref}
\usepackage{caption}

\def\eqx"#1"{{\label{#1}}}
\def\eqn"#1"{{\ref{#1}}}

\makeatletter 
\@addtoreset{equation}{section}
\makeatother  

\def\squarebox#1{\hbox to #1{\hfill\vbox to #1{\vfill}}}
\def\boxit#1{\vbox{\hrule\hbox{\vrule\kern6pt
          \vbox{\kern6pt#1\kern6pt}\kern6pt\vrule}\hrule}}

\newtheorem{thm}{Theorem}

\newcolumntype{L}[1]{>{\raggedright\let\newline\\\arraybackslash\hspace{0pt}}m{#1}}
\newcolumntype{C}[1]{>{\centering\let\newline\\\arraybackslash\hspace{0pt}}m{#1}}
\newcolumntype{R}[1]{>{\raggedleft\let\newline\\\arraybackslash\hspace{0pt}}m{#1}}

\usepackage{booktabs,multirow}
\usepackage{tikz}
\usetikzlibrary{shapes}

\usepackage{threeparttable}
\usepackage{xurl}

\begin{document}

\baselineskip=19pt
\begin{center}
{\Large \bf  Design and Sample Size Determination for Multiple-dose Randomized Phase II Trials for Dose Optimization}
\end{center}
\begin{center}
{\bf Peng Yang$^{1,2}$, Daniel Li$^{3}$, Ruitao Lin$^{2}$, Bo Huang$^{4}$, Ying Yuan$^{2, *}$}
\end{center}

\begin{center}
$^{1}$Department of Statistics, Rice University, Houston, Texas 77005, U.S.A. \\
$^{2}$Department of Biostatistics, The University of Texas MD Anderson Cancer Center\\
Houston, Texas 77030, U.S.A.\\
$^{3}$Bristol Myers Squibb, Seattle, WA, 98109, U.S.A. \\
$^{4}$Pfizer Inc., New York, NY 10017, U.S.A. \\
{*Email: yyuan@mdanderson.org}
\end{center}

\noindent{\textbf{Abstract}} $\quad$
The conventional more-is-better dose selection paradigm, which targets the maximum tolerated dose (MTD), is not suitable for the development of targeted therapies and immunotherapies as the efficacy of these novel therapies may not increase with the dose. The U.S. Food and Drug Administration (FDA) has launched Project Optimus ``to reform the dose optimization and dose selection paradigm in oncology drug development'', and recently published a draft guidance on dose optimization, which outlines various approaches to achieve this goal. One highlighted approach involves conducting a randomized phase II trial following the completion of a phase I trial,  where multiple doses (typically including the MTD and one or two doses lower than the MTD) are compared to identify the optimal dose that maximizes the benefit-risk tradeoff. This paper focuses on the design of such a multiple-dose randomized trial, specifically the determination of the sample size. We propose a MERIT (Multiple-dosE RandomIzed Trial design for dose optimization based on toxicity and efficacy) design that can be easily implemented with pre-calculated decision boundaries included in the protocol. We generalized the standard definitions of type I error and power to accommodate the unique characteristics of dose optimization and derived a decision rule along with an algorithm to determine the optimal sample size. Simulation studies demonstrate that the resulting MERIT design has desirable operating characteristics. To facilitate the implementation of the MERIT design, we provide software, available at \url{www.trialdesign.org}.\\


\baselineskip=22pt

\newpage
\section{Introduction} \label{sec:Introduction}

Conventionally, the primary objective of phase I oncology trials is to identify the maximum tolerated dose (MTD) and subsequently progress it to phase II and III trials to evaluate efficacy. This MTD-centered, more-is-better dose-selection paradigm, established based on cytotoxic chemotherapies, is problematic for novel targeted therapies and immunotherapies \citep{ratain2014redefining, zang2014adaptive, yan2018phase, Ratain2021dose, shah2021drug}. Many of these novel therapies are characterized by shallow dose-response, and as a result, the MTD may not be reached within a clinically active dose range. In addition, efficacy may not monotonically increase with the dose, and often plateaus after the dose reach a certain level \citep{cook2015early, sachs2016optimal}. Therefore, the dose that optimizes the benefit-risk tradeoff may occur at a dose lower than the MTD. Given that over the past several decades, most new oncology drugs are targeted therapies, the U.S. Food and Drug Administration (FDA) Oncology Center of Excellence recently launched Project Optimus ``to reform the dose optimization and dose selection paradigm in oncology drug development'' \citep{FDA2022} and released draft guidance ``Optimizing the Dosage of Human Prescription Drugs and Biological Products for the Treatment of Oncologic Diseases" \citep{FDA2023}. 

According to the FDA's draft guidance, one approach to dose optimization is to conduct a randomized multiple-dose phase II trial after the completion of phase I dose escalation and identification of the MTD. Patients are randomized to two or more doses, typically including the MTD and doses lower than the MTD, to identify the optimal biological dose (OBD)  that maximizes the benefit-risk tradeoff. \cite{Zirkelbach2022} reviewed the dose optimization in FDA initial approvals (2019-2021) of small molecules and antibody-drug conjugate for oncologic indications, including the use of randomized trials with multiple dosages for dose optimization. For example, Belantamab mofodotin, an antibody-drug conjugate, was granted accelerated approval for treating relapsed or refractory multiple myeloma. In the phase I dose escalation trial,  patients received dosages ranging from 0.03 to 4.6 mg/kg IV every 3 weeks, and no MTD was reached. In the subsequent trial, patients were randomized to two doses (2.5 or 3.4mg/kg) to establish the OBD.  Both doses demonstrated similar efficacy, with an objective response rate (ORR) of 31\% for the 2.5 mg arm (97 patients) and 34\% for the 3.4 mg arm (99 patients), but the 2.5 mg arm had a better safety profile, with fewer serious adverse events. Based on the data, the recommended OBD for approval was 2.5 mg.

Multiple-arm randomized trial designs have been proposed to select the optimal treatment from multiple treatments. \cite{dunnett1955multiple} came up with a multiple comparison procedure and compared several treatments with control using a multivariate analogue of student t-distribution to account for the correlation between comparisons. \cite{dunnett1984selection} proposed a single-stage design to identify the best treatment from several under the normal assumption with pooled variance. \cite{simon1985randomized} reviewed the sources of variability influencing the results of phase II trials and found that randomization and selection among new agents or schedules are of value both scientifically and logistically. \cite{whitehead1986sample} developed a Bayesian design to evaluate several treatments and then progress the most promising treatment to compare with a randomized control.  \cite{thall1988two} and \cite{thall1989two} proposed a two-stage design to select the best of several treatments with optimal sample size and decision rules by minimizing the total expected sample size. \cite{schaid1990optimal} developed a two-stage design to identify the best treatment for survival improvement over a standard control.  

The design of multiple-dose randomized trials for dose optimization poses several challenges beyond those of the multiple treatment selection designs discussed previously. First, most aforementioned designs focus on a single efficacy endpoint. For dose optimization, it is imperative to consider both efficacy and toxicity (or more general benefit and risk) in order to fully evaluate the exposure-response relationship and identify the OBD that yields optimal benefit-risk tradeoff. As shown later, considering efficacy and toxicity endpoints jointly is significantly more challenging than considering only a single efficacy endpoint. For example, when considering only efficacy, the null hypothesis is simply that all treatments are not efficacious. In contrast, when considering both toxicity and efficacy, the null hypothesis has various forms (e.g, some doses are efficacious but too toxic and thus not acceptable) that should be accounted for in the trial design. The expanded dimension of endpoints has profound implications on standard design properties such as type I error and power. Generalization of standard type I error and power is needed. Second, the aforementioned designs focus on the selection among several independent treatments (e.g., each arm has a different drug). In contrast, randomized dose optimization trials have the same drug with ordered doses, and it is important to incorporate this feature (e.g., toxicity in the higher dose arm should be no lower than that in the lower dose arm) when constructing the hypothesis and trial design. Third, determining the OBD involves dealing with high-dimensional data, which includes the assessment of a multitude of  characteristics of treatment such as safety, efficacy, biological activities, pharmacokinetics (PK), pharmacodynamics (PD), tolerability, among others. This makes the sample size determination challenging. FDA draft guidance on dose optimization states that ``The trial should be sized to allow for sufficient assessment of activity, safety, and tolerability for each dosage. The trial does not need to be powered to demonstrate statistical superiority of a dosage or statistical non-inferiority among the dosages." \citep{FDA2023}.  The guidance does not provide specific guidance on the sample size determination, a forefront question all investigators will face when designing randomized dose optimization trials. 

To address these challenges, we propose a \underline{m}ultiple-dos\underline{e} \underline{r}andom\underline{i}zed \underline{t}rial (MERIT) design for dose optimization. We proposed a 2-stage OBD selection framework to reduce the dimension of the dose optimization, making the problem statistically tractable. This framework allows us to focus on the identification of OBD admissible dose set based on the primary toxicity and efficacy endpoints, rather than the OBD. To accommodate the distinctive aspects of dose randomization, which necessitate the simultaneous assessment of efficacy and toxicity, we generalize the standard definition of type I error and power. Based on that, we derive the decision rule and provide  an algorithm to determine the sample size.  Our method provides the first rigorous approach to determining the sample size for randomized dose optimization trials. The resulting MERIT is simple to implement, entailing solely a comparison between the observed counts of efficacy and toxicity against predefined decision thresholds, while maintaining rigorous control over type I error and power. 

The remainder of the paper is organized as follows. In Section \ref{sec:Trial_Design}, we propose the MERIT design, including the generalized definition of type I error and power, decision rule, and algorithm to determine decision boundaries. 
 In Section \ref{sec:Simulation}, we conduct comprehensive simulations to assess the operating characteristics of the design and its robustness. We close with a discussion in Section \ref{sec:Discussion}.

\section{Method} \label{sec:Trial_Design}

\subsection{OBD Admissible Set} 
Consider a multiple-dose randomized trial, where a total of $J\times n$ patients are equally randomized to $J$ doses, $d_1 < d_2 < \cdots < d_J$.  In most applications, $J=$ 2 or 3, and the highest dose $d_J$ is often the MTD or maximum administered dose (when the MTD is not reached) identified in the phase I trial. 
Let $Y_T$ and $Y_E$ denote primary endpoints for toxicity and efficacy, respectively. Here, ``toxicity" and ``efficacy" are used generally to represent the potential risks and benefits/activities of the treatment being studied, respectively. We focus on the case that $Y_T$ and $Y_E$ are binary. Examples of $Y_T$ include dose-limiting toxicity (e.g., grade 3 or worse adverse event scored by the Common Terminology Criteria for Adverse Events (CTCAE)), dichotomized total toxicity burden that accounts for different CTCAE grades and types of toxicities, and dose tolerability (i.e., the rate of dose discontinuation/reduction/interruption due to toxicity). Examples of $Y_E$ include objective response and efficacy surrogate endpoints (e.g., PD endpoints and target receptor occupancy).

One major challenge in dose optimization is that although $Y_T$ and $Y_E$ are pivotal factors to consider, the determination of the OBD involves a multitude of factors beyond $Y_T$ and $Y_E$. As highlighted in the FDA's guidance \citep{FDA2023},  the identification of the OBD  is a highly complex process that involves the comprehensive assessment of diverse facets of the treatment, including safety, efficacy, PK, PD, and tolerability. Each of these factors entails the consideration of multiple endpoints. For example, safety assessment involves adverse events of different organs with different severity grades and the rate of treatment discontinuation, treatment duration, and compliance (delay, interruption); efficacy assessment may involve depth of response, duration of response, progression-free survival, and overall survival; PK endpoints often include area under the curve (AUC), Cmin, Cmax, Tmax, and t$_{1/2}$; and PD assessment may involve multiple PD biomarkers. The inherent multi-dimensional complexity of OBD selection renders the formulation of precise OBD selection criteria virtually unattainable. Even if such criteria were possible, their practical implementation would be prohibitively intricate. Now, the challenge is that, in the absence of a well-defined OBD criterion, it is impossible to define the design and investigate its statistical properties.

To render the problem tractable, we introduce a two-stage OBD selection framework, aligning with both established drug development practices and the FDA's guidance:
\begin{itemize}
\item Stage 1: Identify the OBD admissible dose set, denoted as ${\cal A}$, defined as a set of doses that satisfy certain prespecified toxicity and efficacy criteria pertaining to primary endpoints $Y_T$ and $Y_E$.
\item Stage 2: Select the OBD from ${\cal A}$ based on a comprehensive evaluation of collective risk and benefit data, including safety, efficacy, PK, PD, and tolerability. 
\end{itemize}
Our methodology development will primarily focus on controlling the operating characteristics of Stage 1,  particularly the type I error, power and sample size pertaining to the identification of ${\cal A}$.  Because the final OBD selection in Stage 2 is confined within ${\cal A}$, ensuring robust operational characteristics in Stage 1 is imperative. A well-designed Stage 1 often translates into good overall operating characteristics for the dose optimization trial because the majority of uncertainty comes from Stage 1.  Stage 2,  based on collective risk and benefit data, generally yields more precise decisions than Stage 1.

Let $\pi_{T,j} = \Pr(Y_T=1|d_j)$ and $\pi_{E,j}=\Pr(Y_E=1|d_j)$ denote the probability of the occurrence of toxicity and efficacy events, respectively, for dose $d_j$, $j=1, \cdots, J$. As often the case in practice, we assume that $\pi_{T,j}$ and $\pi_{E,j}$ are non-decreasing with respect to the dose, i.e., $\pi_{T,1}  \leq \cdots \leq \pi_{T,J}$ and $\pi_{E,1} \leq \cdots \leq \pi_{E,J}$, while noting that this assumption is accommodated but not required by our design.  Let $\phi_{T,0}$ denote the null toxicity rate that is high and deemed unacceptable, and $\phi_{T,1}$ denote the alternative toxicity rate that is low and deemed acceptable, with $\phi_{T,0} > \phi_{T,1}$. Similarly, let $\phi_{E,0}$ and $\phi_{E,1}$ denote the null and alternative efficacy rates that are deemed unacceptable and acceptable, respectively, with $\phi_{E,0} < \phi_{E,1}$. A dose $d_j$ is defined as OBD admissible if $\pi_{T,j} \leq \phi_{T,0}$ and $\phi_{E,j} \geq \phi_{E,0}$.

\subsection{Global type I error}


For the purpose of dose optimization, we consider the null hypothesis
\begin{center}
$H_0$: None of the doses is OBD admissible.  
\end{center}
The unique feature and challenge here are that there are multiple parameter configurations corresponding to $H_0$. Precisely, $H_0$ consists of a set of $K = \sum_{j = 1}^{J + 1} j$ hypotheses as follows: 
\begin{align*}
     H_0(s,k): & \ \underbrace{\pi_{T,1} = \pi_{T,2} = \dots = \pi_{T,s} =  \phi_{T,1} }_{ \text{acceptable toxicity}} < \underbrace{\pi_{T, s+1} = \dots = \pi_{T,k} = \pi_{T,k+1} = \dots = \pi_{T,J} = \phi_{T,0}}_{ \text{unacceptable toxicity}}; \\
    & \ \underbrace{\pi_{E,1} = \pi_{E,2} = \dots = \pi_{E,s} = \pi_{E,s+1} = \dots = \pi_{E,k}=  \phi_{E,0}}_{ \text{unacceptable efficacy}} < \underbrace{\pi_{E, k+1} = \dots = \pi_{E,J}= \phi_{E,1}}_{\text{acceptable efficacy}},
\end{align*}
where $s,k \in \{0, 1, \dots, J\}$ with $s \leq k$. We use $s = k = 0$ to represent that all doses are efficacious but unacceptably toxic and $s = k = J$ to denote all doses are safe but futile. Of note, $H_0(s,k)$ has incorporated the toxicity and efficacy ordering of doses. 

Given $H_0(s, k)$, the doses are partitioned into three groups. The first group is  $\{d_1, \dots, d_s\}$, where each dose has acceptable toxicity but unacceptable efficacy; the second group is $ \{d_{s+1}, \dots, d_k\}$, where each dose has unacceptable toxicity and unacceptable efficacy; the third group is $ \{d_{k+1}, \dots, d_J\}$, where each dose has acceptable efficacy but unacceptable toxicity. Some of these groups can be empty. Thus, under $H_0(s, k)$, there is no OBD. Table \ref{tbl:Example_null_alt} provides all possible $H_0(s, k)$ for $J=2$ and 3. The existence of multiple parameter configurations under the null hypothesis stems from jointly considering toxicity and efficacy, i.e., a dose can be unacceptable due to unacceptable toxicity or/and unacceptable efficacy. This challenge does not occur in treatment-selection designs such as \cite{thall1988two} and \cite{thall1989two} that consider only efficacy. In these designs, there is only a single simple null hypothesis (i.e., all treatments are not efficacious).

One implication of the existence of multiple null hypothesis configurations is that the standard definition of type I error is not sufficient to fully characterize operating characteristics of randomized dose optimization trials because type I error depends on and differs across $H_0(s, k)$. 
To address this issue, let $\alpha(s, k) = \text{Pr}(\text{reject } H_0(s, k)| H_0(s,k))$ denote the type I error under $H_0(s,k)$. We define {\it global type I error} that encompasses all $H_0(s, k)$ as follows:
\begin{eqnarray}\label{eqn:alpha}
    \begin{aligned}
    \alpha & = \text{Pr}(\text{reject } H_0 | H_0) \\
    & = \underset{0 \leq s \leq J, \,\, s \leq k \leq J}{\max} \{\alpha(s, k)    \}.
    \end{aligned}
\end{eqnarray}
As the global type I error represents the maximum of type I errors across all $H_0(s, k)$, if we control $\alpha$ at a nominal value $\alpha^*$, type I error for each possible $H_0(s, k)$ is controlled strictly below $\alpha^*$. The proposed MERIT design will control the global type I error.

Given the decision rule defined above, the type I error under $H_0(s, k)$ is given by:
\begin{eqnarray}\label{eqn:alpha_sk}
    \begin{aligned}
\alpha(s,k)    & =  \text{Pr}( \text{reject } H_0(s, k)| H_0(s,k)) \\
     & = 1 - \big \{ \big( 1 -  \text{Pr}(n_T \leq m_T, n_E \geq m_E; n, \phi_{T,1}, \phi_{E,0}) \big )^{s} \\
     & ~~~~~~~~~\times \big( 1 - \text{Pr}(n_T \leq m_T, n_E \geq m_E; n, \phi_{T,0}, \phi_{E,0}) \big)^{k - s} \\
     & ~~~~~~~~~\times \big( 1 - \text{Pr}(n_T \leq m_T, n_E \geq m_E; n, \phi_{T,0}, \phi_{E,1}) \big)^{J - k}  \big \},
    \end{aligned}
\end{eqnarray}
where $n_T$ and $n_E$ denote the number of toxicity and efficacy, respectively. By enumerating all possible combinations of $s$ and $k$, $0 \leq s \leq J, s \leq k \leq J$, we obtain the global type I error as defined  by (\ref{eqn:alpha}).

\subsection{Generalized power}

To define the power, we consider the alternative hypothesis:
\begin{equation*}
    H_1: \text{At least one dose is OBD admissible}.
\end{equation*}
Similar to $H_0$, there are multiple parameter configurations corresponding to $H_1$.  Specifically, $H_1$ encompasses a collection of $\sum_{j=1}^J j$ hypotheses as follows:
\begin{align*}
     H_1(u,v): & \ \underbrace{\pi_{T,1} = \pi_{T,2} = \dots = \pi_{T,u} = \pi_{T,u+1} = \dots = \pi_{T,v}=  \phi_{T,1} }_{\text{acceptable toxicity}} < \underbrace{\pi_{T, v+1} = \dots = \pi_{T,J}= \phi_{T,0} }_{ \text{unacceptable toxicity}}; \\
     & \ \underbrace{\pi_{E,1} = \pi_{E,2} = \dots = \pi_{E,u}= \phi_{E,0} }_{ \text{unacceptable efficacy}} < \underbrace{\pi_{E, u+1} = \dots = \pi_{E,v} = \pi_{E,v+1} = \dots = \pi_{E,J}= \phi_{E,1}}_{ \text{acceptable efficacy}},
\end{align*}
where $u, v \in \{0, 1, 2, \dots, J\}$ with $u < v$, with $u = 0$ representing that all doses are efficacious and at least one dose is safe, and $v = J$ representing that all doses are safe and there is at least one dose is efficacious. Given $H_1(u, v)$, the doses are partitioned into three groups. The first group is  $\{d_1, \dots, d_u\}$, where each dose has acceptable toxicity but unacceptable efficacy; the second group is $ \{d_{u+1}, \dots, d_v\}$, where each dose has both acceptable toxicity and efficacy; the third group is $ \{d_{v+1}, \dots, d_J\}$, where each dose has acceptable efficacy but unacceptable toxicity. Thus, under $H_1(u, v)$, there is $v - u$ admissible doses. Table \ref{tbl:Example_null_alt} provides all the possible $H_1(u,v)$ for $J = 2, 3$.

Given $H_1(u,v)$, we could apply the standard definition and define the power of the design as Pr(reject $H_0 | H_1(u,v)$). This standard definition, however, does not sufficiently account for the characteristics of dose optimization. To see this, consider a trial with two doses ($d_1$, $d_2$), where $d_1$ is safe but futile ($\pi_{T, 1} = \phi_{T,1}$ and $\pi_{E, 1} = \phi_{E, 0}$) and $d_2$ is safe and efficacious ($\pi_{T, 2} = \phi_{T,1}$ and $\pi_{E, 2} = \phi_{E, 1}$). The (incorrect) decision that only $d_1$ is OBD admissible (i.e., ${\cal A} =\{d_1\}$) leads to rejecting $H_0$, but excludes the possibility of correctly identifying the OBD in step 2. Thus, it is not appropriate to count such rejection of $H_0$ as power. 

To address this issue, we define two generalized powers: $\beta_1(u,v)$ and $\beta_2(u,v)$, referred to as generalized power I and II, respectively.
\begin{eqnarray*}\label{eqn:beta_1_uv}
    \begin{aligned}
    \beta_1(u,v)  = & \ \text{Pr} (\text{reject } H_0 \text{ and all doses in ${\cal A}$ are truly safe and efficacious}  ~|~ H_1(u,v)) \\
    = & \ \text{Pr} (n_E < m_E | H_1(u,v))^u \big \{ 1 - \big[ 1 - \text{Pr}(n_T \leq m_T, n_E \geq m_E | H_1(u,v)) \big ]^{(v - u)} \big \} \\
    & ~~~~~ \text{Pr} (n_T > m_T | H_1(u,v))^{J - v}.
    \end{aligned}
\end{eqnarray*}

\begin{eqnarray*}\label{eqn:beta_2_uv}
    \begin{aligned}
    \beta_2(u,v)  = & \ \text{Pr} (\text{reject } H_0 \text{ and at least one dose in ${\cal A}$ is truly safe and efficacious}  ~|~ H_1(u,v)) \\
    = & \big \{ 1 - \big[ 1 - \text{Pr}(n_T \leq m_T, n_E \geq m_E | H_1(u,v)) \big ]^{(v - u)} \big \}.
    \end{aligned}
\end{eqnarray*}
Both generalized powers are stricter than the standard power and target the identification of admissible doses. In addition to rejecting $H_0$, $\beta_1(u,v)$ requires that all doses in ${\cal A}$ are truly safe and efficacious, and $\beta_2(u,v)$ requires that at least one dose in ${\cal A}$ is truly safe and efficacious. This additional requirement is consistent with the stepwise decision-making process and ensures the quality of subsequent final OBD selection (i.e., step 2), noting that the final OBD is selected from ${\cal A}$. 

$\beta_1(u,v)$ is stricter than $\beta_2(u,v)$ because the former does not allow any false positive in ${\cal A}$ (i.e., incorrectly identify some futile or/and toxic doses as admissible), while the latter allows that as long as not all doses in ${\cal A}$ are false positive. To achieve the same power, $\beta_1(u,v)$ requires a larger sample size than $\beta_2(u,v)$. The choice of $\beta_1(u,v)$ or $\beta_2(u,v)$ depends on the trial characteristics and the user's tolerability of false positives. In the context of the two-stage decision-making procedure described previously, a false positive is of less concern than standard hypothesis testing because the false positive (made in step 1) could be identified and corrected later in step 2.  Therefore, generalized power II may be a good option when reducing the sample size is of top priority. 

Generalized power I and II do not require all truly admissible doses to be identified and included in ${\cal A}$. That is, it is possible that some of the admissible doses are not selected into ${\cal A}$ (e.g., only one of the two truly admissible doses is selected). It may seem appealing to consider a stricter definition that all truly admissible doses are correctly identified and included in ${\cal A}$. This, however, is excessively stringent for dose optimization trials. These trials are primarily exploratory and are not meant to demonstrate statistical superiority or non-inferiority among doses, as noted in the FDA draft guidance \citep{FDA2023}. Demanding that all truly admissible doses be identified and included in ${\cal A}$ would require an impractically large sample size to achieve a reasonable power such as 80\%. For instance, if there were three doses and a 20\% false positive rate for identifying OBD admissible doses, the power to correctly identify all admissible doses would only be $0.8^3=0.51$. For brevity, we will refer to generalized power I and II as power in the subsequent discussions.

The MERIT design aims to control the power under $H_1$ at a prespecified nominal level $\beta_1^*$ or $\beta_2^*$. The challenge is that $H_1$ encompasses a large collection of $H_1(u,v)$, and each $H_1(u,v)$ leads to a different power. Along a similar line as controlling global type I error, we define global power I and II, denoted as $\beta_1$ and $\beta_2$, as 
\begin{equation} \label{eqn:beta}
    \beta_{i}  = \underset{u, v \in \{0, \cdots, J\}, u<v }{\min} \beta_i (u, v),  ~ \text{for } i = 1,2.
\end{equation}
 We aim to control $\beta_1 \ge \beta_1^*$ or $\beta_2 \ge \beta_2^*$. This task is facilitated by the least favorable set, a subset of configurations of $H_1$ that contains the configuration leading to the lowest power over all $H_1(u,v)$'s. The least favorable set generalizes the idea of the least favorable configuration considered by \cite{thall1988two, thall1989two} and \cite{gibbons1999selecting}, which is applicable to the case that there is a single least favorable configuration. In our case, it is impossible to pinpoint a single least favorable configuration as it depends on the values of $(\phi_{T,0}, \phi_{T,1}, \phi_{E,0}, \phi_{E,1})$. However, it can be shown that there exists a subset of configurations (i.e.,  least favorable set) that will contain the least favorable configuration, as described in Theorem \ref{thm:L-LFC}. 

\begin{thm} \label{thm:L-LFC}
Define the least favorable set $\widetilde{H}_1 = \{ {H}_1(j), j=1, \cdots, J\}$, where
\begin{equation*}
   {H}_1 (j) = 
    \begin{pmatrix}
     \pi_{T,1} = \dots = \pi_{T,j - 1} = \phi_{T,1} & \pi_{T,j} = \phi_{T,1}   & \pi_{T,j+1} = \dots = \pi_{T,J} = \phi_{T,0} \\
     \underbrace{\pi_{E,1} = \dots = \pi_{E,j-1} = \phi_{E,0}}_{\text{safe but futile}} & \underbrace{\pi_{E,j} = \phi_{E,1}}_{\text{safe and efficacious}} & \underbrace{\pi_{E,j+1} = \dots = \pi_{E,J} = \phi_{E,1}}_{\text{toxic and efficacious}} 
    \end{pmatrix}.
\end{equation*}
For any $H_1(u,v)$, with $u, v \in \{0, 1, 2,\dots,J\}$ and $u < v$,  there exists an ${H}_1 (j)$ such that $\beta_i(j)\le \beta_i(u, v)$, $i = 1, 2$, where $\beta_1(j)$ and $\beta_2(j)$ denote the generalized power I and II under ${H}_1 (j)$, respectively. 
 \end{thm}
The proof is provided in the Appendix. Theorem \ref{thm:L-LFC} indicates that to calculate global power, we only need to focus on the least favorable set, consisting of $J$ hypotheses under which only one of the doses is OBD admissible. Therefore, the global power in (\ref{eqn:beta}) can be equivalently defined as
\begin{equation} \label{eqn:beta_GLFC}
    \beta_i  = \underset{j \in 
    \{1,\dots, J\}}{\min} \beta_i (j) ~ \text{for } i = 1,2.
\end{equation}
This property greatly simplifies the calculation of the global power as it reduces the minimization space from $\sum_{j=1}^J j$ to $J$. Under the MERIT design, $\beta_1(j)$ is given by
\begin{eqnarray*}\label{eqn:beta_1_LFC}
    \begin{aligned}
    \beta_1(j)  = & \ \text{Pr} (n_{E,1} < m_E, \dots, n_{E,j-1} < m_E, n_{T,j+1} > m_T, \dots, n_{T,j} > m_T, \\
    & ~~~~~~~~~ n_{E,j} \geq m_E, n_{T,j} \leq m_T ~ |  H_1(j)), \\
    \end{aligned}
\end{eqnarray*}
and $\beta_2(j)$ is given by 
\begin{eqnarray*}\label{eqn:beta_2_LFC}
    \begin{aligned}
    \beta_2(j)  = & \ \text{Pr} (n_{E,j} \geq m_E, n_{T,j} \leq m_T ~ |  H_1(j)), \\
    \end{aligned}
\end{eqnarray*}
where $n_{T,j}$ and $n_{E,j}$ denote the total number of patients who experienced efficacy and toxicity at $d_j$, respectively.

\subsection{MERIT Design} \label{sec:Opt_design}

The MERIT design is simple, described by the following three steps:
\begin{enumerate}
\item[(a)] Specify the target global type I error $\alpha^*$ and global power $\beta_1^*$ or $\beta_2^*$.
\item[(b)] Equally randomize $J\times n$ patients to $d_1, \cdots, d_J$.
\item[(c)] At the end of the trial, in any dose arm $d_j$, $j\in (1, \cdots, J)$, if $n_{E, j} \ge m_E$ and $n_{T, j} \le m_T$, we reject $H_0$ and select the OBD admissible set ${\cal A}$ as the doses satisfying $n_{E, j} \ge m_E$ and $n_{T, j} \le m_T$, where $m_E$ and $m_T$ are decision boundaries.
\end{enumerate}
The design parameters ($n, m_T, m_E$) are determined through numerical search using the following algorithm. This process aims to meet the prescribed criteria for both type I error and power. 
\begin{enumerate}
\item Set $n=1, \cdots, N$, where $N$ is a large number.
\item Given a value of $n$, enumerate all possible values of $m_E, m_T \in (0, 1, \cdots, n)$. Given a set of $(n, m_E, m_T)$, calculate type I error and power based on equations (\ref{eqn:alpha}) and (\ref{eqn:beta_GLFC}), respectively.
\item Repeat steps 1 and 2 until we find the smallest $n$ and corresponding $m_E$ and $m_T$, such that $\alpha\le \alpha^*$ and $\beta_1 \ge \beta_1^*$ or $\beta_2 \ge \beta_2^*$.  
\end{enumerate}
The resulting design is optimal in the sense that it minimizes $n$, given the global type I and global power constraints $\alpha\le \alpha^*$ and $\beta_1 \ge \beta_1^*$ or $\beta_2 \ge \beta_2^*$.

In step 2 of the algorithm, the evaluation of type I errors and power is based on Monte Carlo simulation. Specifically, we simulate $\{ n_{T,j}, n_{E,j} \}_{j = 1}^J$ for all possible null hypotheses and $J$ least favorable alternative hypotheses. For each null or alternative hypothesis, we first simulate latent variables $(X_{T,j}, X_{E,j})$ from a bivariate normal distribution:
\begin{eqnarray*}
\begin{pmatrix}
    X_{T,j} \\ 
     X_{E,j}
\end{pmatrix}
\sim \mathcal{N}
\begin{pmatrix}
    \begin{pmatrix}
        0 \\ 
        0
    \end{pmatrix},
    \begin{pmatrix}
    1 & \rho \\
    \rho & 1 
    \end{pmatrix}
\end{pmatrix},
\end{eqnarray*}
where $\rho$ is a prespecified correlation coefficient and we take $\rho = 0.5$ as default value. Then,  $Y_{T,j}$ and $Y_{E,j}$ are generated as $(Y_{T,j}, Y_{E,j}) = (\mathds{1}\{ \Phi(X_{T,j})  \leq \phi_{T,j} \}, \mathds{1}\{ \Phi(X_{E,j})  \leq \phi_{E,j} \})$, where $\Phi(\cdot)$ is the standard normal CDF. This data generation procedure ensures that $Y_{T,j}$ and $Y_{E, j}$ are correlated and their marginal probabilities are $\phi_{T,j}$ and $\phi_{E,j}$, respectively. The type I error rate and power are evaluated empirically by simulating a large number of replicates of $Y_{T,j}$ and $Y_{E, j}$. To facilitate the application of MERIT, the software to calculate ($n, m_T, m_E$) and simulate the operating characteristics is available at \url{www.trialdesign.org}.

Table \ref{tbl:Design_para} provides the optimal design parameters of $(n, m_T, m_E)$ under some common settings of randomized dose optimization trials with $J = 2, 3$, $\alpha^* = 0.1, 0.2, 0.3$, $\beta_1^* = \beta_2^* = 0.6, 0.7, 0.8$, $\rho = 0.5$, $(\phi_{T,0}, \phi_{T,1}) = (0.4, 0.2)$, and various values of $(\phi_{E, 0}, \phi_{E, 1})$. Depending on the power and other settings, the required sample size ranges from 15 to 60 per arm.  For example, for two-dose randomized trials, given $(\phi_{E, 0}, \phi_{E, 1}) = (0.2, 0.4)$ and $\beta_1^*=0.6$, the required sample size is around 30, 25, and 23 per arm for $\alpha^* = 0.1, 0.2, \text{ and } 0.3$. To reach $\beta_1^* = 0.8$, the required sample size increases to 47, 44, and 44 per arm. Additionally, if we focus on reaching $\beta_2^* = 0.6$, the required sample size is 26, 18, and 18 for $\alpha^* = 0.1, 0.2, \text{ and } 0.3$ and it increases to 45, 35, 24 for $\beta_2^* = 0.8$.

Figure \ref{fig:SV_samp} depicts the relationship between the sample size $n$ and the type I error $\alpha^*$ when power $\beta_1^*$ and $\beta_2^*$ is fixed at 0.6, 0.7, 0.8. As expected, a higher power requires a larger sample size, and also more doses require a larger sample size per arm. However, $\beta_1$ and $\beta_2$ behave differently in terms of how $n$ changes with $\alpha^*$. We first discuss $\beta_1$. Unlike standard power calculation with a single endpoint (e.g., efficacy), where $n$ monotonically decreases with the type I error given a fixed power, here $n$ often plateaus when $\alpha^*$ exceeds a certain value (e.g., $0.3$ for $J = 2$ and $0.2$ for $J = 3$ in our setting), given a fixed $\beta_1^*$. That is, increasing $\alpha^*$ does not necessarily increase power. This unique characteristic of dose optimization trials stems from the consideration of two endpoints and the definition of generalized power $\beta_1$. Relaxing $\alpha^*$  will increase the probability of identifying truly admissible  doses, but at the same time also increase the probability of incorrectly claiming inadmissible doses as admissible. As a result, $\beta_1$ may not increase. To see the point, consider a trial with three doses $d_1$, $d_2$, and $d_3$, where $d_2$ is the OBD admissible and $d_1$ and $d_3$ are not admissible. Increasing $\alpha^*$ will result in a larger $m_T$ and smaller $m_E$. This increases the probability of identifying $d_2$ as admissible, but at the same time, it will also increase the probability of incorrectly identifying $d_1$ and $d_3$ as admissible. As a result, power may not increase. Under this case, the sample size is mostly constrained by $\beta_1^*$. Increasing $\alpha^*$ will not reduce the sample size. In contrast, for $\beta_2$, the relationship between $n$ and $\alpha^*$ is more in line with the standard power calculation --- a larger $\alpha^*$ results in a smaller $n$. The different behavior between $\beta_1$ and $\beta_2$ is due to that $\beta_2$ allows false positives. In the above example, increasing the probability of incorrectly identifying $d_1$ and $d_3$ as admissible will not impact $\beta_2$.

\subsection{Practical consideration for trial implementation} \label{subsec:trial_implementation}
The MERIT design is simple to implement. We elicit unacceptable/acceptable toxicity rate and efficacy rate ($\phi_{T, 0}$, $\phi_{T,1}$, $\phi_{E, 0}$, and $\phi_{E,1}$) from subject matter experts and specify target $\alpha^*$ and $\beta_1^*$ or $\beta_2^*$, and then use the software provided to determine optimal $(m_E, m_T, n)$. Given $(m_E, m_T, n)$, the implementation of MERIT only involves a simple comparison of the observed number of toxicity and efficacy (i.e., $n_{T, j}$ and $n_{E, j}$) with $m_T$ and $m_E$ for each dose arm to determine the OBD admissible set ${\cal A}$. Based on the totality of benefit-risk data, one dose is selected from ${\cal A}$ as the OBD.

Due to the randomness of the small sample size, one issue that may arise in practice is that $n_{T, j}$ may be higher in a lower dose than a higher dose and exceeds $m_T$, and as a result, the lower dose is not admissible due to toxicity but a higher dose is admissible. This violates the monotonicity assumption of  toxicity. We can address this issue by applying the isotonic transformation to $\{n_{T, j}\}$ before applying the design decision rule. Specifically, we apply the PAVA algorithm \cite{brunk1972statistical} to the observed toxicity responses $\{  n_{T, j}, \,\, j=1, \cdots, J \}$, resulting in isotonically transformed toxicity responses $\{ \tilde{n}_{T, j}\}$. We then use $\{ \tilde{n}_{T, j} \}$ to replace ${n}_{T, j}$ to make the decision and determine the OBD admissible dose set. The same procedure can be applied to $\{n_{E, j}\}$ when it is desirable to impose the monotonicity assumption on efficacy. All the results in this article assume the monotonicity assumption on both toxicity and efficacy.

Another practical consideration is that in some trials, it may be desirable to add futility and safety interim monitoring rule to early stop the dose arm that is excessively toxic or/and futile. We incorporate Bayesian rules for conducting interim analyses during the trial:
\begin{itemize}
\item stop arm $j$ for safety if $\Pr(\pi_{T, j} > \phi_{T, 1} | \text{data}) > C_T$,
\item stop arm $j$ for futility if $\Pr(\pi_{E, j} < \phi_{E, 1} | \text{data}) > C_E$,
\end{itemize} 
where $C_T$ and $C_E$ are probability cutoffs (e.g., $C_T=0.95$ and $C_E=0.95$) that are calibrated by simulation such that the stopping probability is reasonably high when the dose is toxic or/and futile and low when the dose is safe and efficacious. The posterior probabilities used in the monitoring rule can be easily calculated using the standard beta-binomial model. Assuming a vague beta prior $\pi_{T, j}, \pi_{E, j}\sim beta(a, b)$, where $a$ and $b$ are small values such as $a=b=0.1$, and let $n^{(t)}$, $n_{T, j}^{(t)}$ and $n_{E, j}^{(t)}$ denote the sample size, the number of toxicity, and the number of efficacy, respectively,  at the $t$th interim. The posteriors of $\pi_{T, j}$ and $\pi_{T, j}$ arise as $\pi_{T, j} | n^{(t)}, n_{T, j}^{(t)} \sim Beta(a+n_{T, j}^{(t)}, b+n^{(t)}-n_{T, j}^{(t)})$ and as $\pi_{E, j} | n^{(t)}, n_{E, j}^{(t)} \sim Beta(a+n_{E, j}^{(t)}, b+n^{(t)}-n_{E, j}^{(t)})$. 

The simulation study presented later demonstrates that adding toxicity and futility monitoring can reduce the average sample size in some scenarios. However, the decision to include toxicity and/or futility monitoring, and if included, how often to conduct the interim monitoring, should be made based on trial characteristics and logistical considerations. For instance, in cases where the toxicity endpoint involves the assessment of tolerability over multiple treatment cycles and/or efficacy takes a long time to evaluate, investigators may opt for no interim monitoring due to the logistical difficulty of halting enrollment for interim analysis. This approach is often acceptable because the doses have already undergone a dose escalation study without significant safety concerns, and the sample size per dose arm is typically small (e.g., 20-30) under MERIT.

In cases where toxicity and futility monitoring is feasible and appropriate, one or two interims may be sufficient and provide a good balance between performance and logistics. The frequency and timing of toxicity and efficacy monitoring do not necessarily have to be the same. It should be noted that adding interim stopping can change the operating characteristics of the design, e.g., reducing power and type I error. Nevertheless, the impact of these changes is usually minor (see the simulation in Section 3), and may be acceptable for dose optimization trials that are not intended to be confirmatory in nature.


\section{Simulation study} \label{sec:Simulation}

We conducted simulation studies to evaluate the operating characteristics of MERIT design. We considered common settings with $J = 2, 3$, $\alpha^* = 0.1, 0.2, 0.3$, $\beta_1^* = \beta_2^* = 0.8$, $(\phi_{T,0}, \phi_{T,1}) = (0.4, 0.2)$, and $(\phi_{E, 0}, \phi_{E, 1}) = (0.1, 0.3)$, (0.2, 0.4), (0.3, 0.5) and (0.4, 0.6). For each setting, we considered 16 scenarios of null hypotheses (see Table \ref{tbl:Example_null_alt}) to assess type I error rate, and 9 scenarios representing all possible alternative hypotheses (see Table \ref{tbl:Example_null_alt}) to assess power. We simulated data based on the latent variable approach as described in Section \ref{sec:Opt_design} with $\rho=0.5$. In what follows, we describe simulation results for $(\phi_{E, 0}, \phi_{E, 1}) = (0.2, 0.4)$. The results for $(\phi_{E, 0}, \phi_{E, 1}) = (0.1, 0.3)$, (0.3, 0.5), and (0.4, 0.6) are generally similar and provided in Supplementary Materials. 

\begin{figure}[H]
    \centering
    \includegraphics[width=0.8\textwidth]{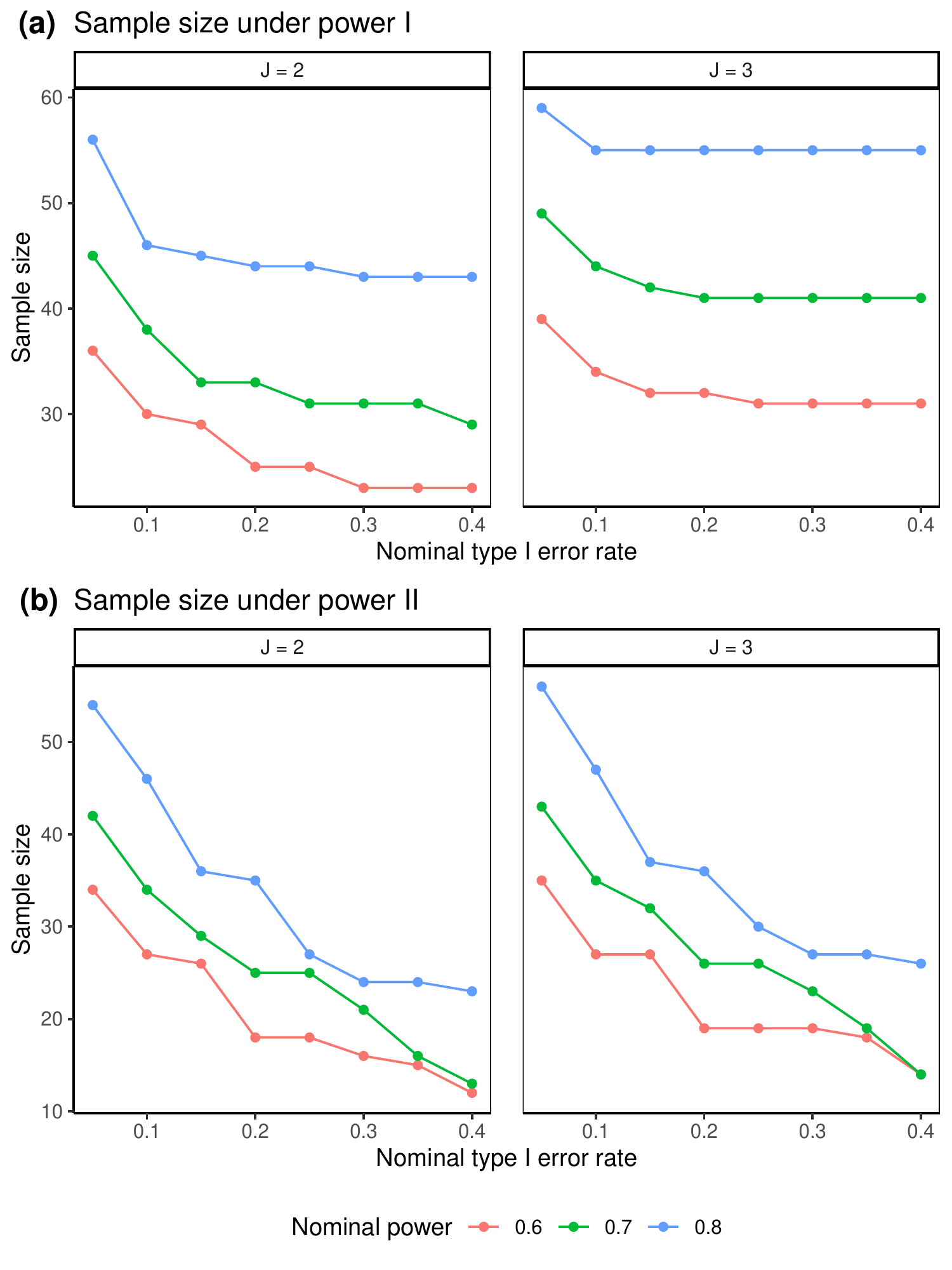}
    \caption{Sample size under different type I error rate and power I (panel (a)) and power II (panel (b)) when $(\phi_{T,0}, \phi_{T,1}) = (0.4, 0.2)$ and $(\phi_{E,0}, \phi_{E,1}) = (0.2, 0.4)$. }
    \label{fig:SV_samp}
\end{figure}

Figure \ref{fig:OC_main} (a) and (b) display the type I error rate $\alpha$ and power $\beta_1$. MERIT is able to control $\alpha$ at its nominal level across all 16 scenarios. In some scenarios (e.g., scenarios 2, 3, 5, 8, 9, 10, 12, 13, 15), $\alpha$ is well below the nominal value. This is because MERIT controls global type I error (i.e., the worst case across all null hypotheses) and thus is expected to be conservative in some null scenarios. One may wonder why in some settings, e.g., when $J=3$ and $\alpha^*=0.2$ or 0.3, the largest $\alpha$ is still substantially lower than the nominal value. The reason is that in these settings, the power is the limiting factor, as explained in Section \ref{sec:Opt_design} and Figure \ref{fig:SV_samp} (a). In these specific settings, in order to satisfy the (global) power constraint, the resulting decision boundaries happen to lead to conservative type I error. Across all alternative hypothesis scenarios, MERIT controls $\beta_1$ at or above the nominal value of 80\% (Figure \ref{fig:OC_main} (b)). For scenarios 17, 19, 20, 23, and 25, which belong to the least favorable set, $\beta_1$ is close to 80\%. In other scenarios, $\beta_1$ tends to be higher than 80\% because by design MERIT controls the global power (i.e., power of the worst scenario) at the nominal value. 

Figure \ref{fig:OC_main} (c) and (d) shows the type I error rate $\alpha$  and power $\beta_2$. MERIT is able to control $\alpha$ and $\beta_2$ at the nominal levels across all scenarios. Compared to that under $\beta_1$, $\alpha$ under $\beta_2$ is closer to the nominal value. This is because under $\beta_2$, $n$ does not plateau with $\alpha$ (see Figure \ref{fig:SV_samp} (b)), and we can adjust $n$ to make $\alpha$ close to its nominal value.

\begin{figure}[H]
    \centering
    \includegraphics[width=0.9\textwidth]{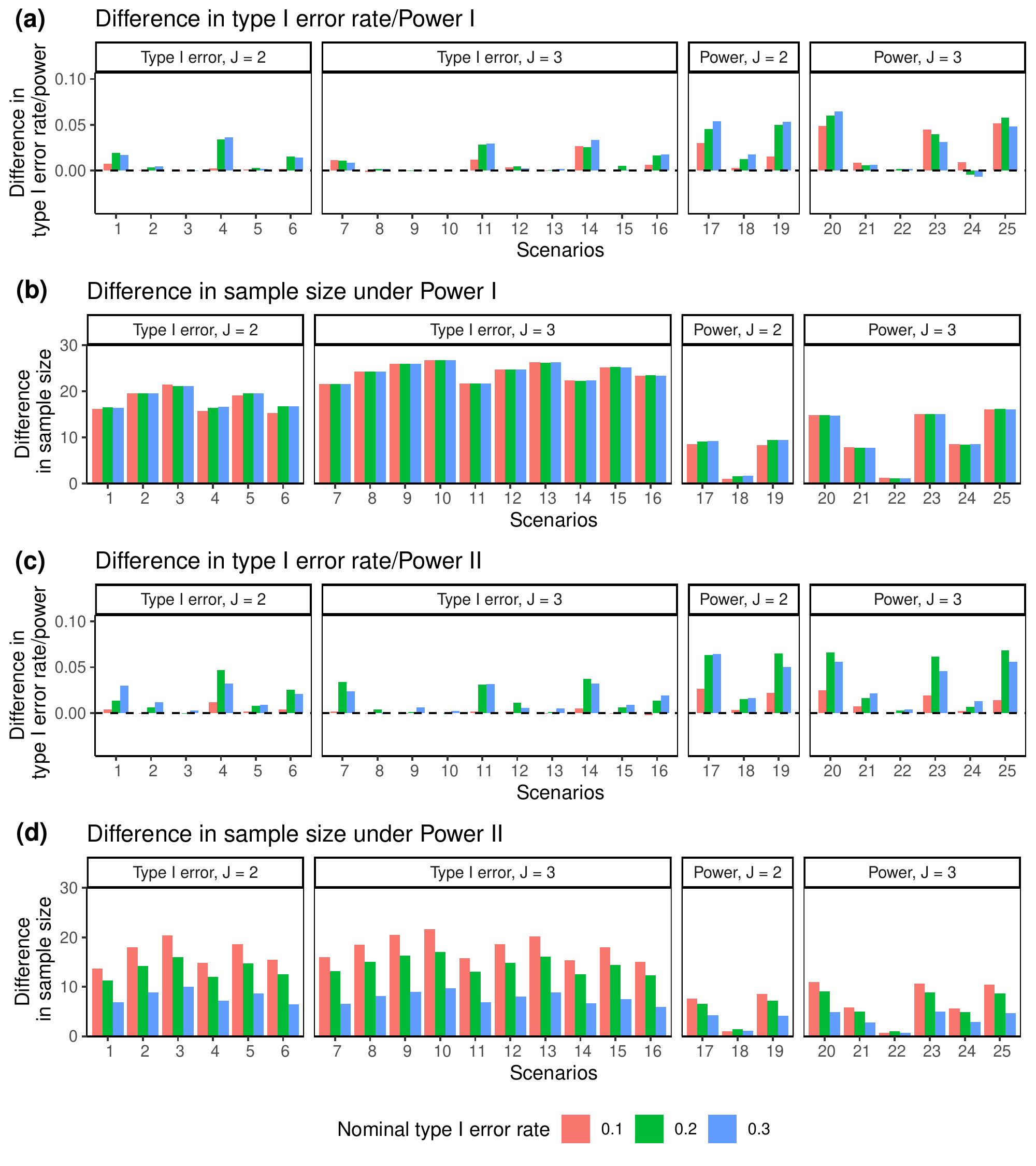}
    \caption{The difference in type I error/power and sample size between without and with an interim toxicity and futility monitoring when $(\phi_{T,0}, \phi_{T,1}) = (0.4, 0.2)$, $(\phi_{E,0}, \phi_{E,1}) = (0.2, 0.4)$, and $\alpha^*=0.1, 0.2$ or 0.3. Panels (a) and (b) show the results for $\beta_1^*=0.8$, and panels (c) and (d) show the results for $\beta_2^*=0.8$. }
    \label{fig:IT_main}
\end{figure}

We further evaluated the operating characteristics of MERIT when interim toxicity and futility monitoring is added to drop overly toxic or futile dose arms. We considered the case with one interim analysis performed when half of the patients are enrolled, and the case with two interim analyses performed when one- and two-thirds of the patients are enrolled. We employed the interim monitoring rule described in Section \ref{subsec:trial_implementation} with $C_T = C_E = 0.95$. Figure \ref{fig:IT_main} shows the results with one interim analysis, and the results with two interim analyses are similar and provided in Supplementary Materials. The results show that adding interim monitoring yields sizable sample size saving in most scenarios under null hypotheses and some scenarios under alternative hypotheses when dose arms are toxic (e.g., scenarios 17, 20, 21, 23) or futile (e.g., scenarios 19, 23, 24, 25). Interim monitoring often leads to slightly more conservative type I error and slightly reduced power (e.g., the reduction is often $<0.05$). Therefore, we recommend adding interim monitoring when it is logistically feasible.

The determination of MERIT design parameters requires specification of the correlation between $Y_T$ and $Y_E$ (i.e., $\rho$). We evaluated the sensitivity of the MERIT design with respect to $\rho$. We used $\rho=0.5$ to design MERIT, and simulated data using $\rho = 0.25$ and $0.75$. The results show that MERIT is robust to the misspecification of $\rho$, see Supplementary Materials for details.

\section{Discussion} \label{sec:Discussion}
We have proposed the MERIT design for dose optimization. MERIT controls type I error and power while optimizing the sample size. One advantage of MERIT is its simplicity. Implementing MERIT only involves a simple comparison of the observed number of toxicity and efficacy with prespecified decision boundaries. To obtain these decision boundaries, we only need to elicit unacceptable/acceptable toxicity rate and efficacy rate $(\phi_{T,0}, \phi_{T,1}, \phi_{E,0}, \phi_{E,1})$ from subject matter experts and specify target type I error $\alpha^*$ and power $\beta_1^*$ or $\beta_2^*$. The software provided can be easily used to determine the design decision boundaries, which can be included in the trial protocol. 

MERIT assumes that toxicity and efficacy endpoints are binary. A possible extension is to accommodate other types of endpoints, such as ordinal, continuous,  or time-to-event endpoints. In addition, MERIT focuses on phase II trials. There is a significant amount of literature on seamless phase II-III designs \citep{stallard2011seamless} that merges phase II and III and enables the combination of data from both phases for more efficient inference.  We may apply the phase II-III design framework to MERIT to further streamline drug development, shorten timelines, and enhance design efficiency. 

\section*{Supplementary Material}
The Supplementary Material contains additional simulation results.

\bigskip
\bigskip
\begin{center}
{\Large \bf Appendix} \label{sec:Appendix}
\end{center}
\bigskip

 \noindent {\large 1. Proof of Theorem \ref{thm:L-LFC}} \\

\noindent Let $S_{1} = \{1, 2, \dots, u \}$, $S_{2} =  \{u+1, \dots, v\}$ and $S_{3} = \{ v + 1, \dots, J \}$ denote the index of futile, OBD admissible, and toxic doses, respectively, where $u, v \in \{0, 1, 2, \dots, J\}$ and $u < v$, so that $|S_{2}| \geq 1$ to represent there exists at least one OBD admissible in this set. If $j = 0$ or $v = J$, then $S_{1}$ or $S_{3}$ are empty sets.  Let $\pi = \{ (\pi_{E,1}, \pi_{T,1}), (\pi_{E,2}, \pi_{T,2}), \dots, (\pi_{E,J}, \pi_{T,J}) \}$ denote the set of toxicity and efficacy rate pairs. By assumptions, these underlying true rates follow:
\begin{eqnarray*}
    \begin{aligned}
        \pi_{T,1} &  \leq \pi_{T,2} \leq \cdots \leq \pi_{T,v} \leq \phi_{T,1} \ \ \ \ 
    \phi_{T,0} \leq \pi_{T,v+1} \leq \cdots \leq \pi_{T,J-1} \leq \pi_{T,J} \\ 
    \pi_{E,1}  &  \leq \pi_{E,2} \leq \cdots \leq \pi_{E,u} \leq \phi_{E,0} \ \ \ \ 
    \phi_{E,1} \leq \pi_{E,u+1} \leq \cdots \leq \pi_{E,J-1} \leq \pi_{E,J}. 
    \end{aligned}
\end{eqnarray*}
The two generalized power function is defined as 
\begin{eqnarray*}
    \begin{aligned}
        \beta_1(\pi) & = \Pr(n_{E,1} < m_E, \dots, n_{E,u} < m_E ~ | ~ \pi) \times \Pr(n_{T,v+1} > m_T, \dots, n_{T,J} > m_T ~ | ~ \pi) \times \\
        \times  &  \big \{ 1 - \big ( 1 - \text{Pr} (n_{T,u+1} \leq m_T, n_{E,u+1} \geq m_E ~ | ~ \pi)  \big ) \times \cdots \\
       & ~~~~~ \times \big ( 1 - \text{Pr} (n_{T,v} \leq m_T, n_{E,v} \geq m_E ~ | ~ \pi)  \big )  \big \},  \\
       \beta_2(\pi)  &  =  \big \{ 1 - \big ( 1 - \text{Pr} (n_{T,u+1} \leq m_T, n_{E,u+1} \geq m_E ~ | ~ \pi)  \big ) \times \cdots \\
       & ~~~~~ \times \big ( 1 - \text{Pr} (n_{T,v} \leq m_T, n_{E,v} \geq m_E ~ | ~ \pi)  \big )  \big \}, 
    \end{aligned}
\end{eqnarray*}
where the $\beta_1(\pi)$ represents the probability of rejecting all the doses from set $S_{1} \cup S_{3}$ and accepting at least one dose from $S_{2}$, and $\beta_2(\pi)$ simply represents the probability of accepting at least one dose from $S_{2}$. Note that $\{n_{E, g} < m_E \}$, $\{n_{T, l} > m_T\}$, and $\{ n_{T,j} \leq m_T, n_{E,j} \geq m_E\}^c$ are stochastically decreasing in each element of $\{ \pi_{E,g}, g \in S_{1}\}$,  $\{ \pi_{T,l}, l \in S_{3}\}$, and $\{ \pi_{E,j}, \pi_{T,j}, j \in S_{2}\}$. As a result, both $\beta_1(\pi)$ and $\beta_2(\pi)$ will be minimized locally when $v - u = 1$, which is equivalent to there existing only one OBD admissible, under the condition that $ (\pi_{T,1}, \pi_{E,1}) = \dots = (\pi_{T,j-1}, \pi_{E,j-1}) =  (\phi_{T,1}, \phi_{E,0})$, $ (\pi_{T,j}, \pi_{E,j}) =  (\phi_{T,1}, \phi_{E,1})$, and $(\pi_{T,j+1}, \pi_{E,j+1}) = \dots = (\pi_{T,J}, \pi_{E,J}) =  (\phi_{T,0}, \phi_{E,1})$, for some $j \in \{1, 2, \dots, J\}$.

\clearpage
\bibliographystyle{plainnat}


\newpage


\begin{table}[h!]
\centering
\caption{Possible null and alternative hypotheses for $J=2$ or 3 doses.}
\begin{tabular}{@{\extracolsep{4pt}}ccccc|cccc@{}}  
\hline\hline
&  &   &   \multicolumn{4}{c}{Null hypotheses}   &   &   \\
\hline
   \multicolumn{5}{c|}{$J = 2$} &   \multicolumn{4}{c}{$J = 3$}       \\
\hline
&  & Scenarios & $d_1$ & $d_2$ & $d_1$ & $d_2$ & $d_3$ & Scenarios \\
\hline
\multirow{2}{*}{$H_0(0, 0)$} & $\pi_{T, j}$ & \multirow{2}{*}{1} & $\phi_{T,0}$ & $\phi_{T,0}$ & $\phi_{T,0}$ & $\phi_{T,0}$  & $\phi_{T,0}$  & \multirow{2}{*}{7} \\ 
& $\pi_{E, j}$ & & $\phi_{E,1}$ & $\phi_{E,1}$ & $\phi_{E,1}$ & $\phi_{E,1}$ & $\phi_{E,1}$  \\ [0.5ex]
\hline
\multirow{2}{*}{$H_0(0, 1)$} & $\pi_{T, j}$ & \multirow{2}{*}{2} & $\phi_{T,0}$ & $\phi_{T,0}$ & $\phi_{T,0}$ & $\phi_{T,0}$  & $\phi_{T,0}$ & \multirow{2}{*}{8} \\ 
& $\pi_{E,j}$ & & $\phi_{E,0}$ & $\phi_{E,1}$ & $\phi_{E,0}$ & $\phi_{E,1}$  & $\phi_{E,1}$  \\ [0.5ex]
\hline
\multirow{2}{*}{$H_0(0, 2)$} &$\pi_{T, j}$ & \multirow{2}{*}{3} &  $\phi_{T,0}$ & $\phi_{T,0}$ & $\phi_{T,0}$ & $\phi_{T,0}$  & $\phi_{T,0}$  & \multirow{2}{*}{9} \\ 
& $\pi_{E, j}$ & & $\phi_{E,0}$ & $\phi_{E,0}$ & $\phi_{E,0}$ & $\phi_{E,0}$ & $\phi_{E,1}$  \\ [0.5ex]
\hline
\multirow{2}{*}{$H_0(0, 3)$} &$\pi_{T, j}$ & \multirow{2}{*}{} &    &   & $\phi_{T,0}$ & $\phi_{T,0}$  & $\phi_{T,0}$  & \multirow{2}{*}{10} \\ 
&$\pi_{E, j}$ & &  &  & $\phi_{E,0}$ & $\phi_{E,0}$ & $\phi_{E,0}$   \\ [0.5ex]
\hline
\multirow{2}{*}{$H_0(1, 1)$} &$\pi_{T, j}$ & \multirow{2}{*}{4} & $\phi_{T,1}$ & $\phi_{T,0}$ & $\phi_{T,1}$ & $\phi_{T,0}$ & $\phi_{T,0}$  & \multirow{2}{*}{11} \\ 
&$\pi_{E, j}$ & & $\phi_{E,0}$ & $\phi_{E,1}$ & $\phi_{E,0}$ & $\phi_{E,1}$ & $\phi_{E,1}$  \\ [0.5ex]
\hline
\multirow{2}{*}{$H_0(1, 2)$} &$\pi_{T, j}$ & \multirow{2}{*}{5} & $\phi_{T,1}$ & $\phi_{T,0}$ & $\phi_{T,1}$ & $\phi_{T,0}$ & $\phi_{T,0}$ & \multirow{2}{*}{12} \\
&$\pi_{E, j}$ & & $\phi_{E,0}$ & $\phi_{E,0}$ & $\phi_{E,0}$ & $\phi_{E,0}$ & $\phi_{E,1}$  \\ [0.5ex]
\hline
\multirow{2}{*}{$H_0(1, 3)$} & $\pi_{T, j}$ & \multirow{2}{*}{ } &   &   & $\phi_{T,1}$ & $\phi_{T,0}$ & $\phi_{T,0}$  & \multirow{2}{*}{13} \\ 
&$\pi_{E, j}$ & &   &   & $\phi_{E,0}$ & $\phi_{E,0}$ & $\phi_{E,0}$   \\ [0.5ex]
\hline
\multirow{2}{*}{$H_0(2, 2)$} &$\pi_{T, j}$ & \multirow{2}{*}{6} & $\phi_{T,1}$ & $\phi_{T,1}$ & $\phi_{T,1}$ & $\phi_{T,1}$ & $\phi_{T,0}$ & \multirow{2}{*}{14}  \\ 
&$\pi_{E, j}$ & & $\phi_{E,0}$ & $\phi_{E,0}$ & $\phi_{E,0}$ & $\phi_{E,0}$ & $\phi_{E,1}$ \\ [0.5ex]
\hline
\multirow{2}{*}{$H_0(2, 3)$} & $\pi_{T, j}$ & \multirow{2}{*}{} &   &   & $\phi_{T,1}$ & $\phi_{T,1}$ & $\phi_{T,0}$ & \multirow{2}{*}{15}  \\
&$\pi_{E, j}$ & &  &  & $\phi_{E,0}$ & $\phi_{E,0}$ & $\phi_{E,0}$   \\ [0.5ex]
\hline
\multirow{2}{*}{$H_0(3, 3)$}  & $\pi_{T, j}$ & \multirow{2}{*}{} &   &   & $\phi_{T,1}$ & $\phi_{T,1}$ & $\phi_{T,1}$  & \multirow{2}{*}{16} \\
&$\pi_{E, j}$  & &  &  & $\phi_{E,0}$ & $\phi_{E,0}$ & $\phi_{E,0}$   \\ [0.5ex]
\hline 
&  &   &   \multicolumn{4}{c}{Alternative hypotheses}   &   &   \\
\hline
\multirow{2}{*}{$H_1(0, 1)$} &$\pi_{T, j}$ & \multirow{2}{*}{17} & $\phi_{T,1}^*$ & $\phi_{T,0}$ & $\phi_{T,1}^*$ & $\phi_{T,0}$  & $\phi_{T,0}$ & \multirow{2}{*}{20} \\ 
&$\pi_{E, j}$ & & $\phi_{E,1}$ & $\phi_{E,1}$ & $\phi_{E,1}$ & $\phi_{E,1}$  & $\phi_{E,1}$  \\ [0.5ex]
\hline
\multirow{2}{*}{$H_1(0, 2)$} &$\pi_{T, j}$ & \multirow{2}{*}{18}  & $\phi_{T,1}^*$ & $\phi_{T,1}^*$  & $\phi_{T,1}^*$ & $\phi_{T,1}^*$  & $\phi_{T,0}$  & \multirow{2}{*}{21} \\ 
&$\pi_{E, j}$ & & $\phi_{E,1}$ & $\phi_{E,1}$ & $\phi_{E,1}$ & $\phi_{E,1}$ & $\phi_{E,1}$  \\ [0.5ex]
\hline
\multirow{2}{*}{$H_1(0, 3)$} &$\pi_{T, j}$ & \multirow{2}{*}{ } &    &   & $\phi_{T,1}^*$ & $\phi_{T,1}^*$  & $\phi_{T,1}^*$  & \multirow{2}{*}{22} \\ 
&$\pi_{E, j}$ & &  &  & $\phi_{E,1}$ & $\phi_{E,1}$ & $\phi_{E,1}$   \\ [0.5ex]
\hline
\multirow{2}{*}{$H_1(1, 2)$} &$\pi_{T, j}$ & \multirow{2}{*}{19} & $\phi_{T,1}$ & $\phi_{T,1}^*$ & $\phi_{T,1}$ & $\phi_{T,1}^*$ & $\phi_{T,0}$  & \multirow{2}{*}{23} \\ 
&$\pi_{E, j}$ & & $\phi_{E,0}$ & $\phi_{E,1}$ & $\phi_{E,0}$ & $\phi_{E,1}$ & $\phi_{E,1}$  \\ [0.5ex]
\hline
\multirow{2}{*}{$H_1(1, 3)$} &$\pi_{T, j}$ & \multirow{2}{*}{ } &   &   & $\phi_{T,1}$ & $\phi_{T,1}^*$ & $\phi_{T,1}^*$  & \multirow{2}{*}{24} \\ 
&$\pi_{E, j}$ & &   &   & $\phi_{E,0}$ & $\phi_{E,1}$ & $\phi_{E,1}$   \\ [0.5ex]
\hline
\multirow{2}{*}{$H_1(2, 3)$} &$\pi_{T, j}$ & \multirow{2}{*}{ } &   &   & $\phi_{T,1}$ & $\phi_{T,1}$ & $\phi_{T,1}^*$  & \multirow{2}{*}{25} \\
&$\pi_{E, j}$ & &  &  & $\phi_{E,0}$ & $\phi_{E,0}$ & $\phi_{E,1}$   \\ [0.5ex]
\hline
\hline
\end{tabular} \\
    ~ Note: $^*$ denotes the doses that are OBD admissible.
\label{tbl:Example_null_alt}
\end{table}

\clearpage
\begin{landscape}
\begin{table}[H]
\begin{center}
\caption{Optimal design parameters when $(\phi_{T,0}, \phi_{T,1}) = (0.4, 0.2)$.} \label{tbl:Design_para}
\begin{tabular}{ccccccccccccc|ccccccccc}
\hline \hline
\multirow{3}{*}{$\phi_{E,0}$} & \multirow{3}{*}{$\phi_{E,1}$} & \multirow{3}{*}{$J$} & \multirow{3}{*}{$\beta^*$} & \multicolumn{9}{c|}{$\beta_1$}                                                                         & \multicolumn{9}{c}{$\beta_2$}                                                                         \\
\cline{5-22}
                        &                          &                    &                        & \multicolumn{3}{c}{$\alpha^* = 0.1$} & \multicolumn{3}{c}{$\alpha^* = 0.2$} & \multicolumn{3}{c|}{$\alpha^* = 0.3$} & \multicolumn{3}{c}{$\alpha^* = 0.1$} & \multicolumn{3}{c}{$\alpha^* = 0.2$} & \multicolumn{3}{c}{$\alpha^* = 0.3$} \\
\cline{5-22}
                        &                          &                    &                        & $n$       & $m_T$       & $m_E$      & $n$       & $m_T$       & $m_E$      & $n$       & $m_T$       & $m_E$      & $n$       & $m_T$       & $m_E$      & $n$       & $m_T$       & $m_E$      & $n$       & $m_T$       & $m_E$      \\
\hline
\multicolumn{2}{c}{\multirow{6}{*}{0.1 ~~ 0.3}}       &                    & 0.6                    & 26       & 7         & 6        & 23       & 6         & 5        & 21       & 6         & 4        & 25       & 6         & 5        & 18       & 5         & 4        & 13       & 4         & 3        \\
\multicolumn{2}{c}{}                               & 2                  & 0.7                    & 33       & 9         & 7        & 30       & 8         & 6        & 27       & 8         & 5        & 33       & 8         & 6        & 24       & 7         & 5        & 19       & 6         & 4        \\
\multicolumn{2}{c}{}                               &                    & 0.8                    & 44       & 12        & 8        & 39       & 11        & 7        & 39       & 11        & 7        & 39       & 11        & 8        & 30       & 8         & 5        & 25       & 7         & 4        \\
\cline{3-22}
\multicolumn{2}{c}{}                               &                    & 0.6                    & 33       & 8         & 6        & 28       & 8         & 6        & 27       & 8         & 5        & 27       & 7         & 6        & 18       & 5         & 4        & 14       & 4         & 3        \\
\multicolumn{2}{c}{}                               & 3                  & 0.7                    & 40       & 11        & 8        & 35       & 10        & 7        & 35       & 10        & 7        & 33       & 9         & 7        & 25       & 7         & 5        & 20       & 6         & 4        \\
\multicolumn{2}{c}{}                               &                    & 0.8                    & 47       & 13        & 9        & 47       & 13        & 9        & 47       & 13        & 9        & 40       & 11        & 8        & 31       & 9         & 6        & 26       & 8         & 5        \\
\hline
\multicolumn{2}{c}{\multirow{6}{*}{0.2 ~~ 0.4}}       &                    & 0.6                    & 30       & 8         & 10       & 25       & 7         & 8        & 23       & 7         & 7        & 26       & 7         & 9        & 18       & 5         & 6        & 18       & 5         & 6        \\
\multicolumn{2}{c}{}                               & 2                  & 0.7                    & 38       & 10        & 12       & 33       & 9         & 10       & 31       & 9         & 9        & 34       & 9         & 11       & 25       & 7         & 8        & 20       & 6         & 6        \\
\multicolumn{2}{c}{}                               &                    & 0.8                    & 47       & 13        & 14       & 44       & 13        & 13       & 44       & 13        & 13       & 45       & 12        & 14       & 35       & 10        & 10       & 24       & 7         & 7        \\
\cline{3-22}
\multicolumn{2}{c}{}                               &                    & 0.6                    & 34       & 9         & 11       & 32       & 9         & 10       & 31       & 9         & 9        & 27       & 7         & 9        & 19       & 5         & 6        & 18       & 5         & 6        \\
\multicolumn{2}{c}{}                               & 3                  & 0.7                    & 44       & 12        & 14       & 41       & 12        & 12       & 41       & 12        & 12       & 36       & 10        & 12       & 26       & 7         & 8        & 23       & 7         & 7        \\
\multicolumn{2}{c}{}                               &                    & 0.8                    & 55       & 16        & 17       & 55       & 16        & 16       & 55       & 16        & 16       & 47       & 13        & 15       & 37       & 11        & 11       & 24       & 7         & 7        \\
\hline
\multicolumn{2}{c}{\multirow{6}{*}{0.3 ~~ 0.5}}       &                    & 0.6                    & 30       & 8         & 13       & 28       & 8         & 12       & 25       & 7         & 10       & 28       & 7         & 12       & 19       & 5         & 8        & 14       & 4         & 6        \\
\multicolumn{2}{c}{}                               & 2                  & 0.7                    & 40       & 11        & 17       & 34       & 10        & 14       & 33       & 10        & 13       & 37       & 10        & 16       & 28       & 8         & 12       & 22       & 6         & 9        \\
\multicolumn{2}{c}{}                               &                    & 0.8                    & 53       & 15        & 22       & 48       & 14        & 19       & 46       & 14        & 18       & 44       & 12        & 18       & 34       & 10        & 14       & 28       & 8         & 11       \\
\cline{3-22}
\multicolumn{2}{c}{}                               &                    & 0.6                    & 37       & 10        & 16       & 34       & 9         & 14       & 34       & 10        & 13       & 34       & 9         & 15       & 19       & 5         & 8        & 19       & 5         & 8        \\
\multicolumn{2}{c}{}                               & 3                  & 0.7                    & 47       & 13        & 20       & 44       & 13        & 18       & 44       & 13        & 17       & 38       & 10        & 16       & 29       & 8         & 12       & 24       & 7         & 10       \\
\multicolumn{2}{c}{}                               &                    & 0.8                    & 57       & 16        & 23       & 57       & 16        & 23       & 57       & 16        & 23       & 49       & 13        & 20       & 35       & 10        & 14       & 28       & 8         & 11       \\
\hline
\multicolumn{2}{c}{\multirow{6}{*}{0.4 ~~ 0.6}}       &                    & 0.6                    & 34       & 9         & 18       & 25       & 7         & 13       & 24       & 7         & 12       & 28       & 7         & 15       & 19       & 5         & 10       & 16       & 4         & 8        \\
\multicolumn{2}{c}{}                               & 2                  & 0.7                    & 43       & 12        & 23       & 35       & 10        & 18       & 34       & 10        & 17       & 38       & 10        & 20       & 25       & 7         & 13       & 18       & 5         & 9        \\
\multicolumn{2}{c}{}                               &                    & 0.8                    & 52       & 15        & 27       & 50       & 15        & 25       & 49       & 15        & 24       & 46       & 13        & 24       & 32       & 9         & 16       & 29       & 8         & 14       \\
\cline{3-22}
\multicolumn{2}{c}{}                               &                    & 0.6                    & 38       & 10        & 20       & 35       & 10        & 18       & 34       & 10        & 17       & 32       & 8         & 17       & 23       & 6         & 12       & 17       & 5         & 9        \\
\multicolumn{2}{c}{}                               & 3                  & 0.7                    & 46       & 13        & 24       & 44       & 13        & 22       & 44       & 13        & 22       & 39       & 11        & 21       & 29       & 8         & 15       & 22       & 6         & 11       \\
\multicolumn{2}{c}{}                               &                    & 0.8                    & 59       & 17        & 30       & 59       & 17        & 30       & 59       & 17        & 30       & 46       & 13        & 24       & 36       & 10        & 18       & 29       & 8         & 14   \\
\hline \hline
\end{tabular}
\end{center}
\end{table}
\end{landscape}

\begin{figure}[H]
    \centering
    \includegraphics[width=0.9\textwidth]{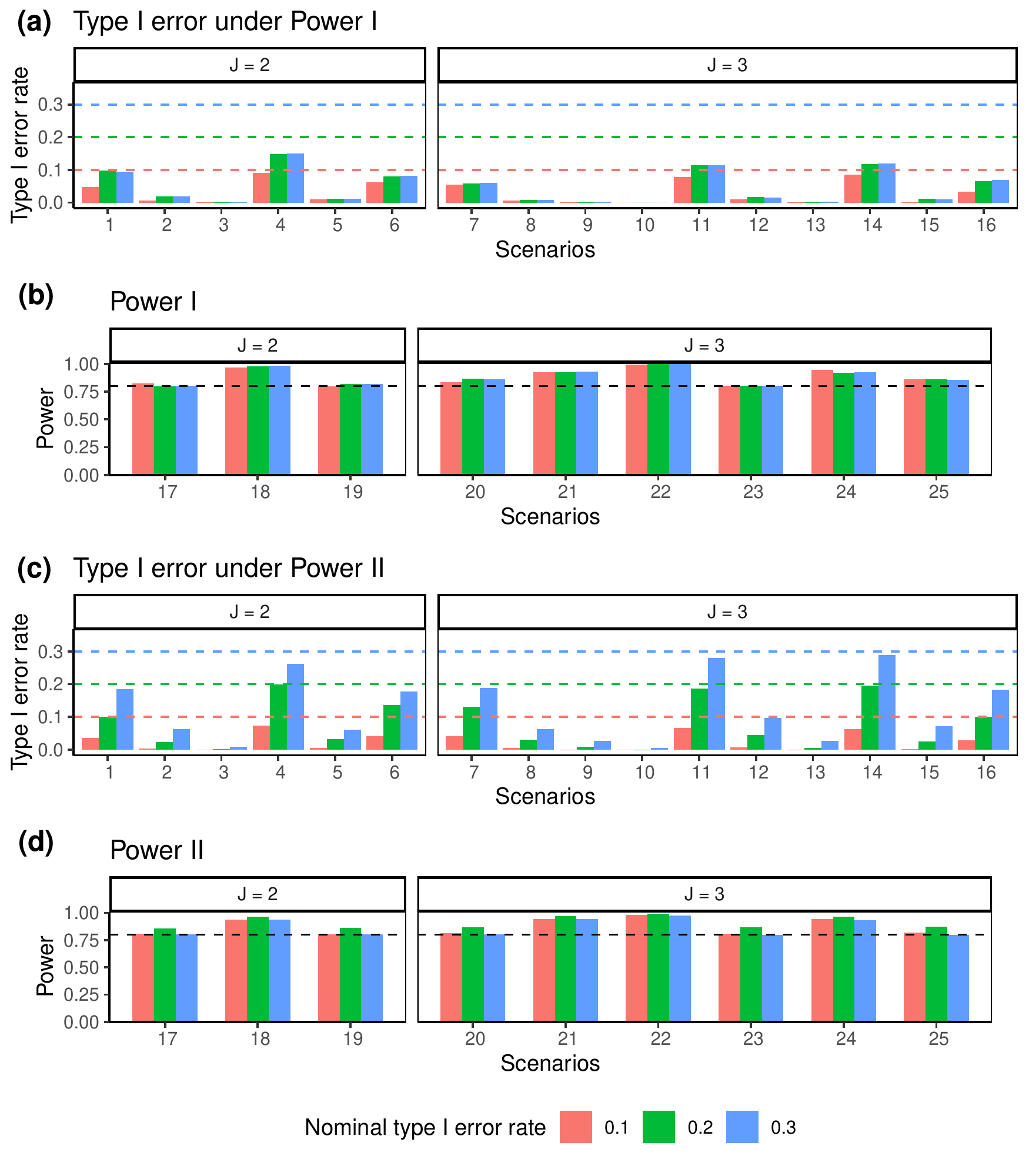}
    \caption{Type I error and power of MERIT design when $(\phi_{T,0}, \phi_{T,1}) = (0.4, 0.2)$, $(\phi_{E,0}, \phi_{E,1}) = (0.2, 0.4)$, and $\alpha^*=0.1, 0.2$ or 0.3. Panels (a) and (b) show the results for $\beta_1^*=0.8$, and panels (c) and (d) show the results for $\beta_2^*=0.8$. The horizontal dashed lines represent the nominal values of type I error and power. The toxicity and efficacy probabilities of scenarios 1-25 are given in Table \ref{tbl:Example_null_alt}. }
    \label{fig:OC_main}
\end{figure}

\section*{Supplementary Material}


\begin{figure}[H]
    \centering
    \includegraphics[width=0.9\textwidth]{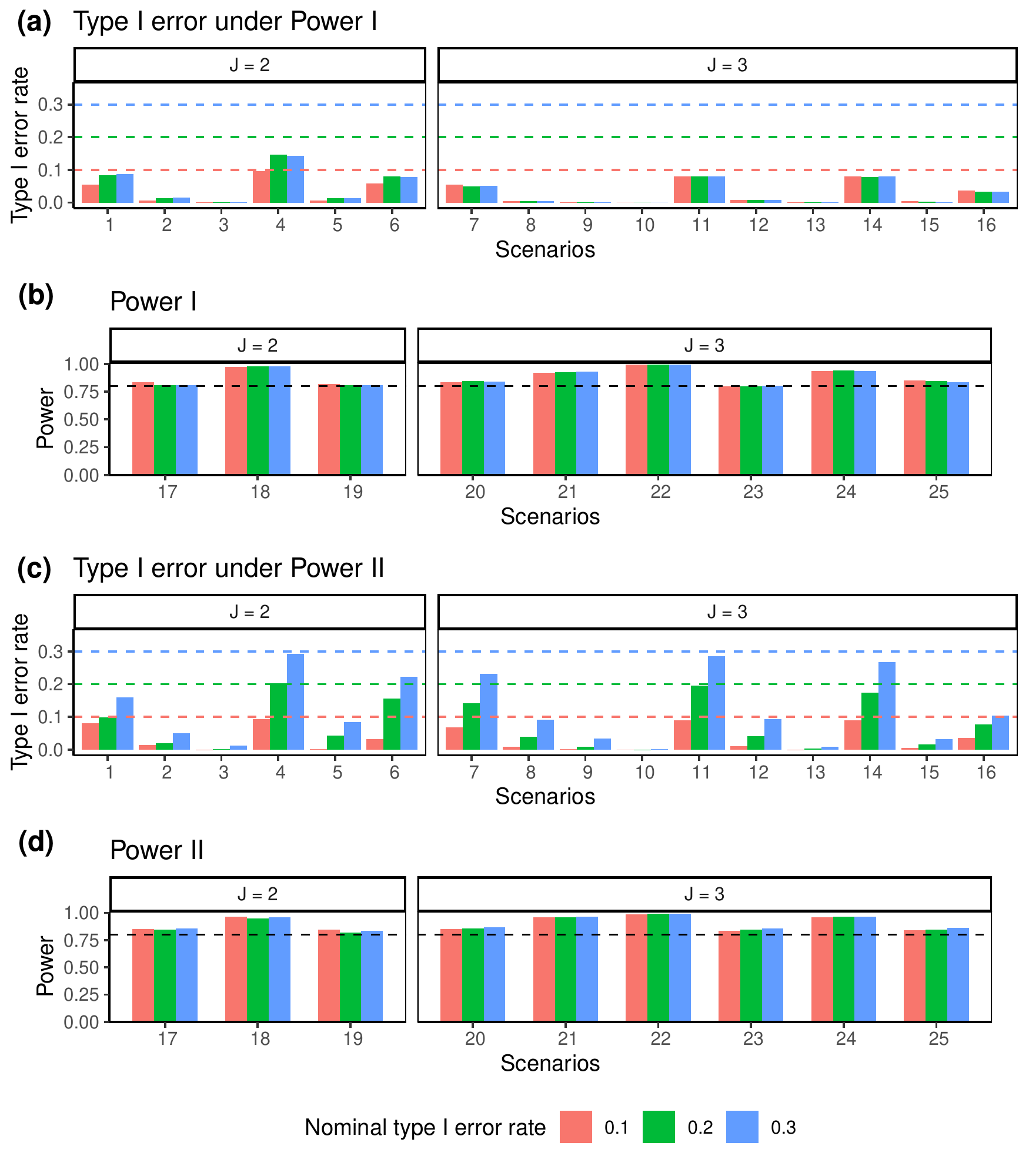}
    \caption{Type I error and power of MERIT design when $(\phi_{T,0}, \phi_{T,1}) = (0.4, 0.2)$, $(\phi_{E,0}, \phi_{E,1}) = (0.1, 0.3)$, and $\alpha^*=0.1, 0.2$ or 0.3. Panels (a) and (b) show the results for $\beta_1^*=0.8$, and panels (c) and (d) show the results for $\beta_2^*=0.8$. The horizontal dashed lines represent the nominal values of type I error and power. The toxicity and efficacy probabilities of scenarios 1-25 are given in Table 1. }
    \label{fig:OC_supp_0.1}
\end{figure}

\begin{figure}[H]
    \centering
    \includegraphics[width=0.9\textwidth]{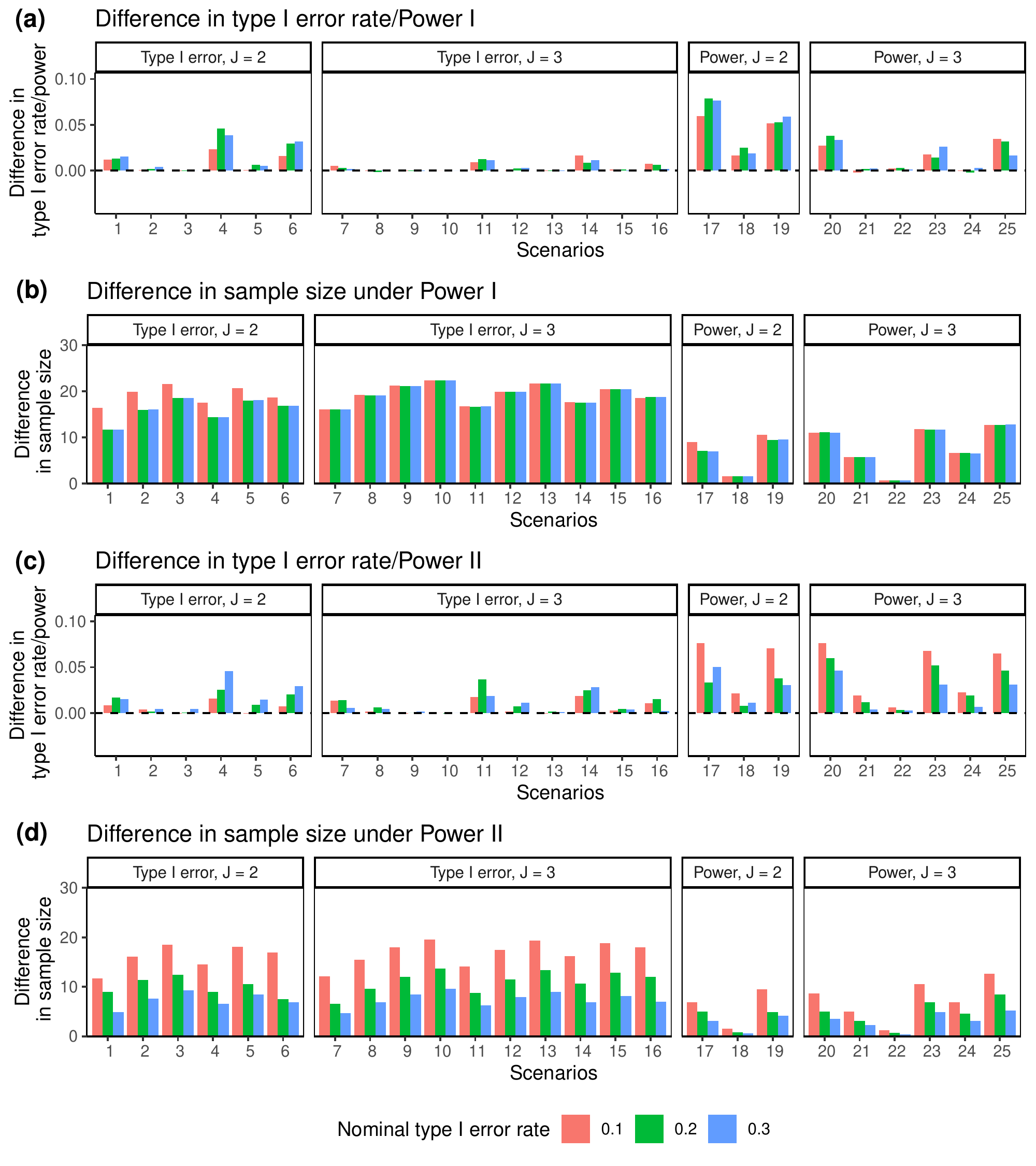}
    \caption{The difference in type I error/power and sample size between without and with an interim toxicity and futility monitoring when $(\phi_{T,0}, \phi_{T,1}) = (0.4, 0.2)$, $(\phi_{E,0}, \phi_{E,1}) = (0.1, 0.3)$, and $\alpha^*=0.1, 0.2$ or 0.3. Panels (a) and (b) show the results for $\beta_1^*=0.8$, and panels (c) and (d) show the results for $\beta_2^*=0.8$. }
    \label{fig:IT_supp_0.1}
\end{figure}


\begin{figure}[H]
    \centering
    \includegraphics[width=0.9\textwidth]{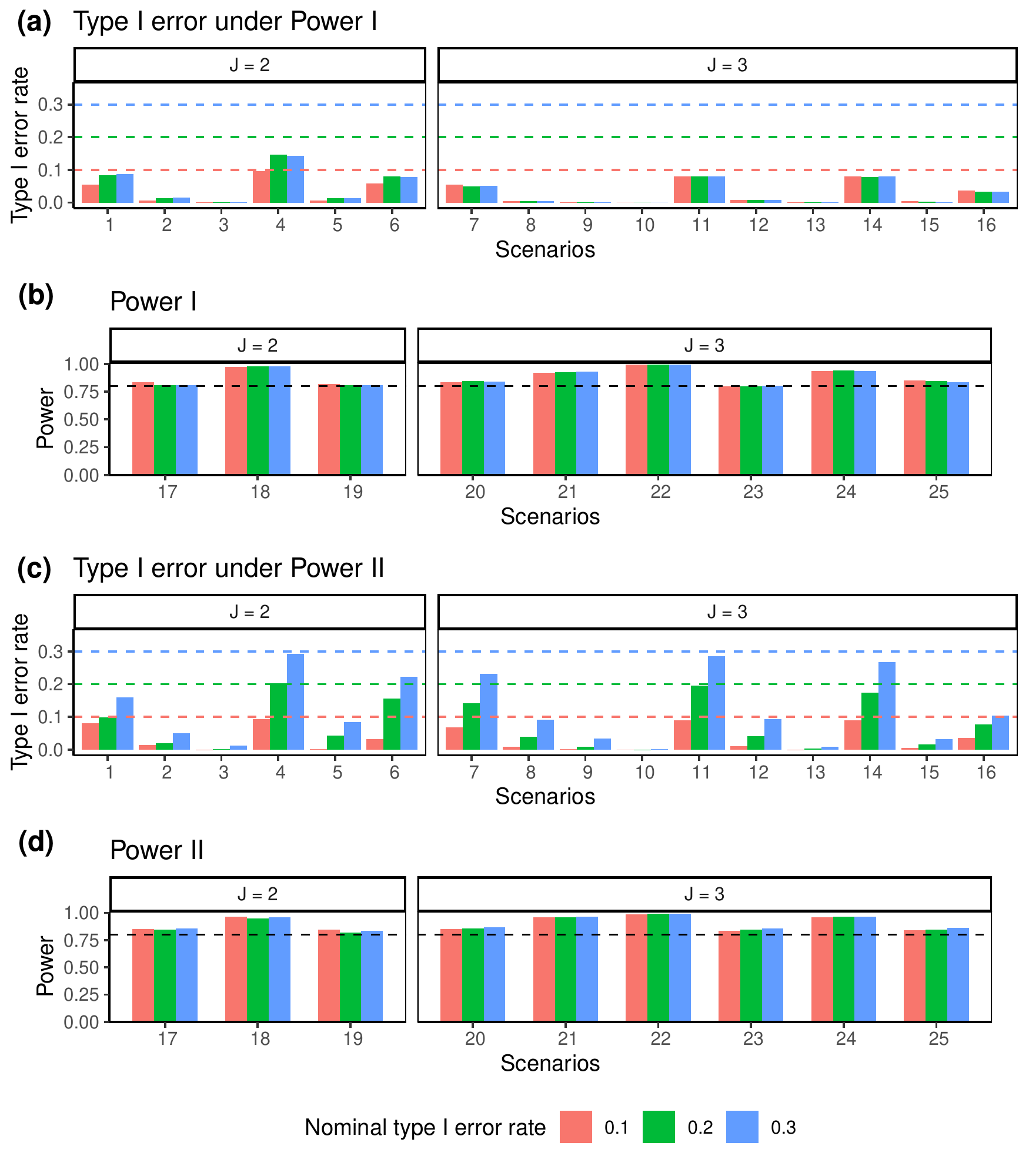}
    \caption{Type I error and power of MERIT design when $(\phi_{T,0}, \phi_{T,1}) = (0.4, 0.2)$, $(\phi_{E,0}, \phi_{E,1}) = (0.3, 0.5)$, and $\alpha^*=0.1, 0.2$ or 0.3. Panels (a) and (b) show the results for $\beta_1^*=0.8$, and panels (c) and (d) show the results for $\beta_2^*=0.8$. The horizontal dashed lines represent the nominal values of type I error and power. The toxicity and efficacy probabilities of scenarios 1-25 are given in Table 1. }
    \label{fig:OC_supp_0.3}
\end{figure}

\begin{figure}[H]
    \centering
    \includegraphics[width=0.9\textwidth]{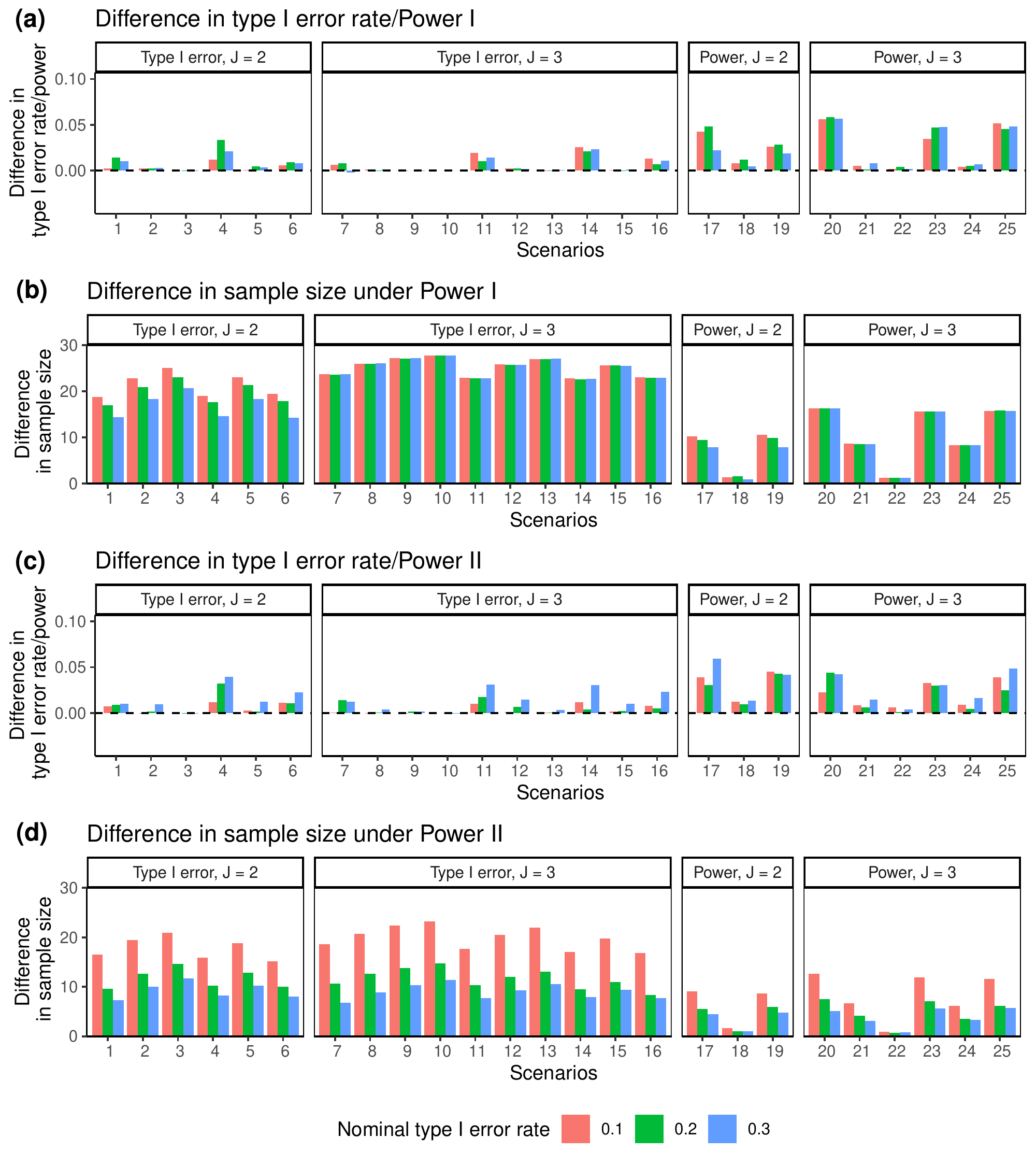}
    \caption{The difference in type I error/power and sample size between without and with an interim toxicity and futility monitoring when $(\phi_{T,0}, \phi_{T,1}) = (0.4, 0.2)$, $(\phi_{E,0}, \phi_{E,1}) = (0.3, 0.5)$, and $\alpha^*=0.1, 0.2$ or 0.3. Panels (a) and (b) show the results for $\beta_1^*=0.8$, and panels (c) and (d) show the results for $\beta_2^*=0.8$. }
    \label{fig:IT_supp_0.3}
\end{figure}


\begin{figure}[H]
    \centering
    \includegraphics[width=0.9\textwidth]{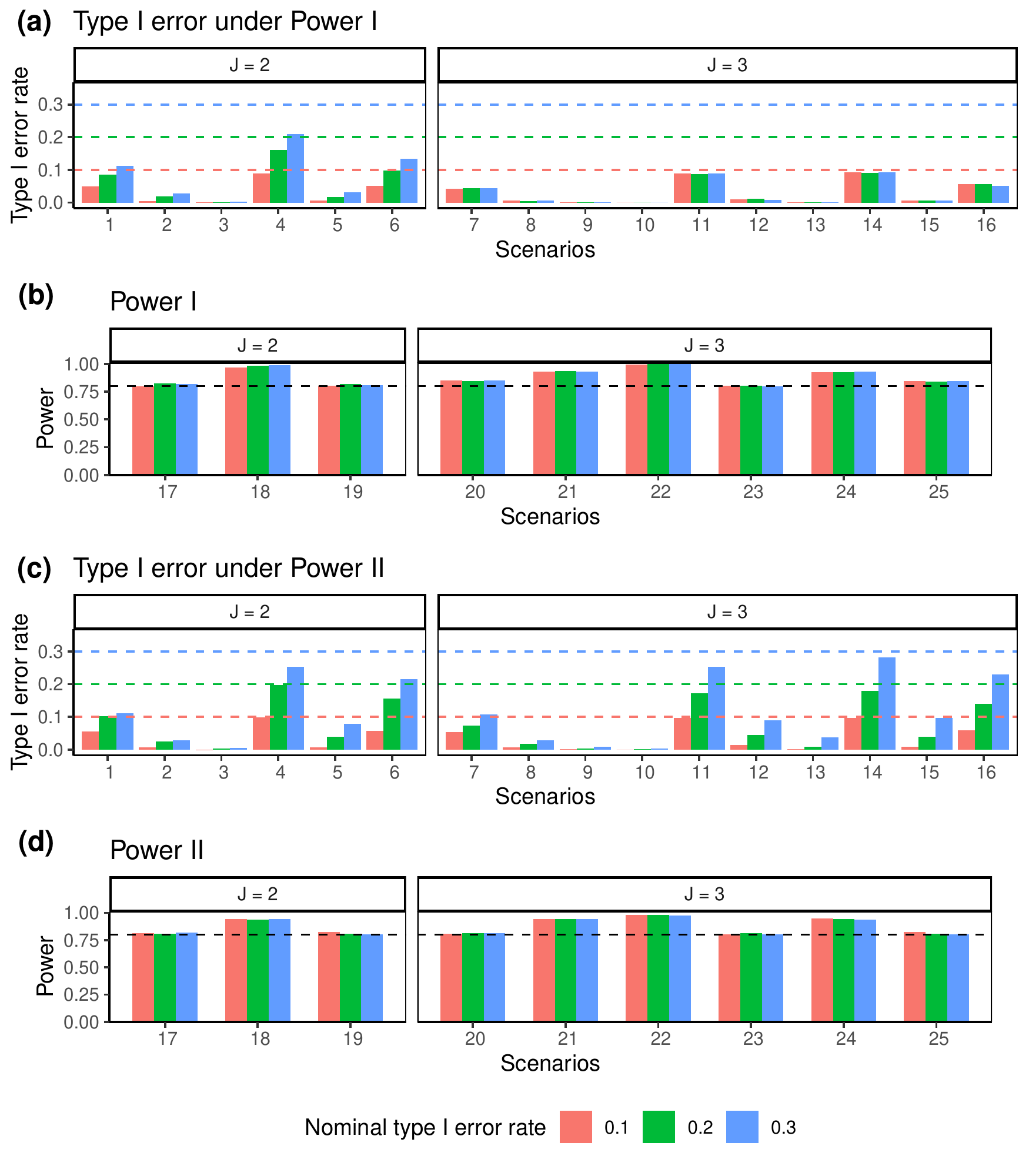}
    \caption{Type I error and power of MERIT design when $(\phi_{T,0}, \phi_{T,1}) = (0.4, 0.2)$, $(\phi_{E,0}, \phi_{E,1}) = (0.4, 0.6)$, and $\alpha^*=0.1, 0.2$ or 0.3. Panels (a) and (b) show the results for $\beta_1^*=0.8$, and panels (c) and (d) show the results for $\beta_2^*=0.8$. The horizontal dashed lines represent the nominal values of type I error and power. The toxicity and efficacy probabilities of scenarios 1-25 are given in Table 1. }
    \label{fig:OC_supp_0.4}
\end{figure}

\begin{figure}[H]
    \centering
    \includegraphics[width=0.9\textwidth]{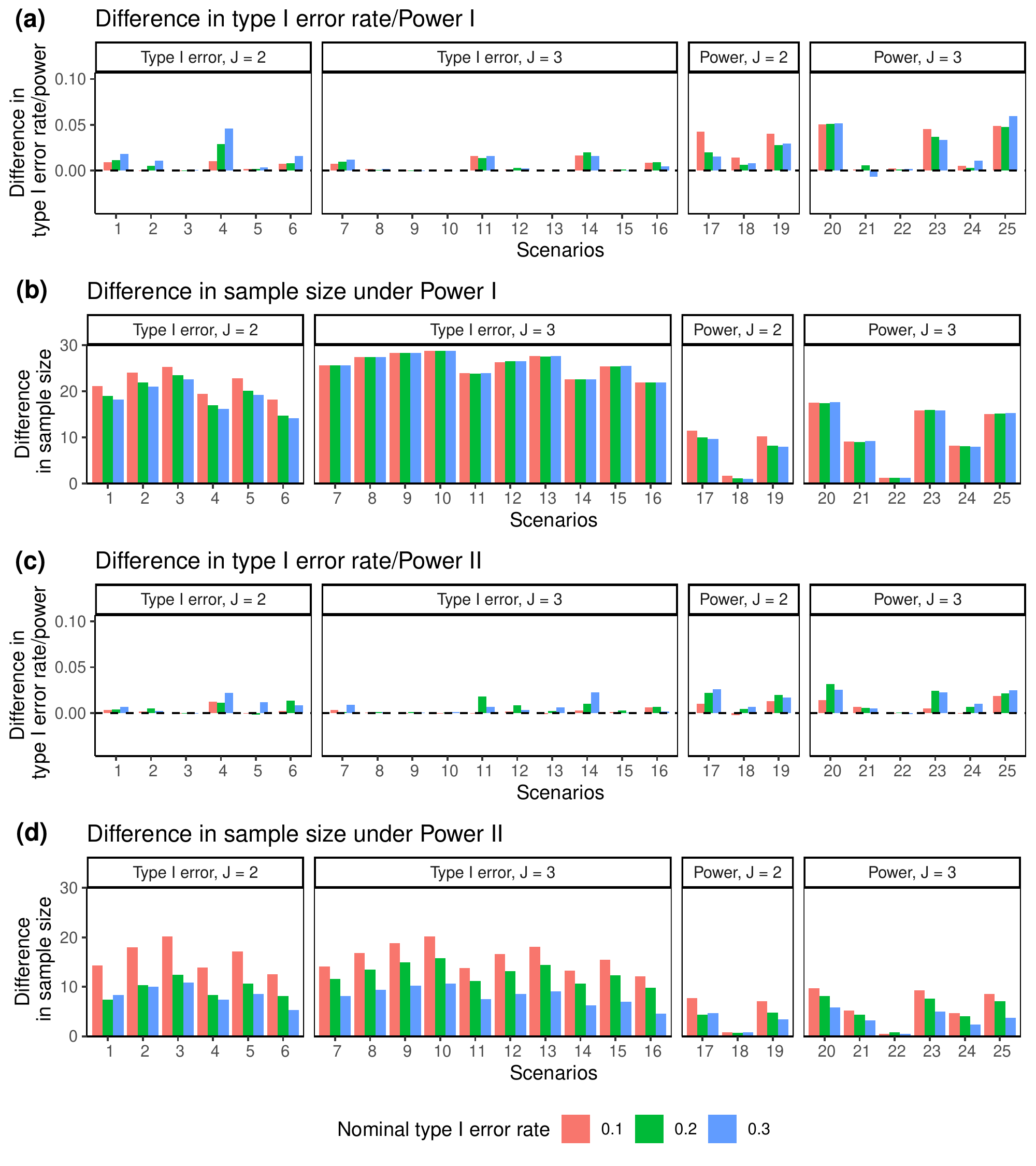}
    \caption{The difference in type I error/power and sample size between without and with an interim toxicity and futility monitoring when $(\phi_{T,0}, \phi_{T,1}) = (0.4, 0.2)$, $(\phi_{E,0}, \phi_{E,1}) = (0.4, 0.6)$, and $\alpha^*=0.1, 0.2$ or 0.3. Panels (a) and (b) show the results for $\beta_1^*=0.8$, and panels (c) and (d) show the results for $\beta_2^*=0.8$. }
    \label{fig:IT_supp_0.4}
\end{figure}

\begin{figure}[H]
    \centering
    \includegraphics[width=0.9\textwidth]{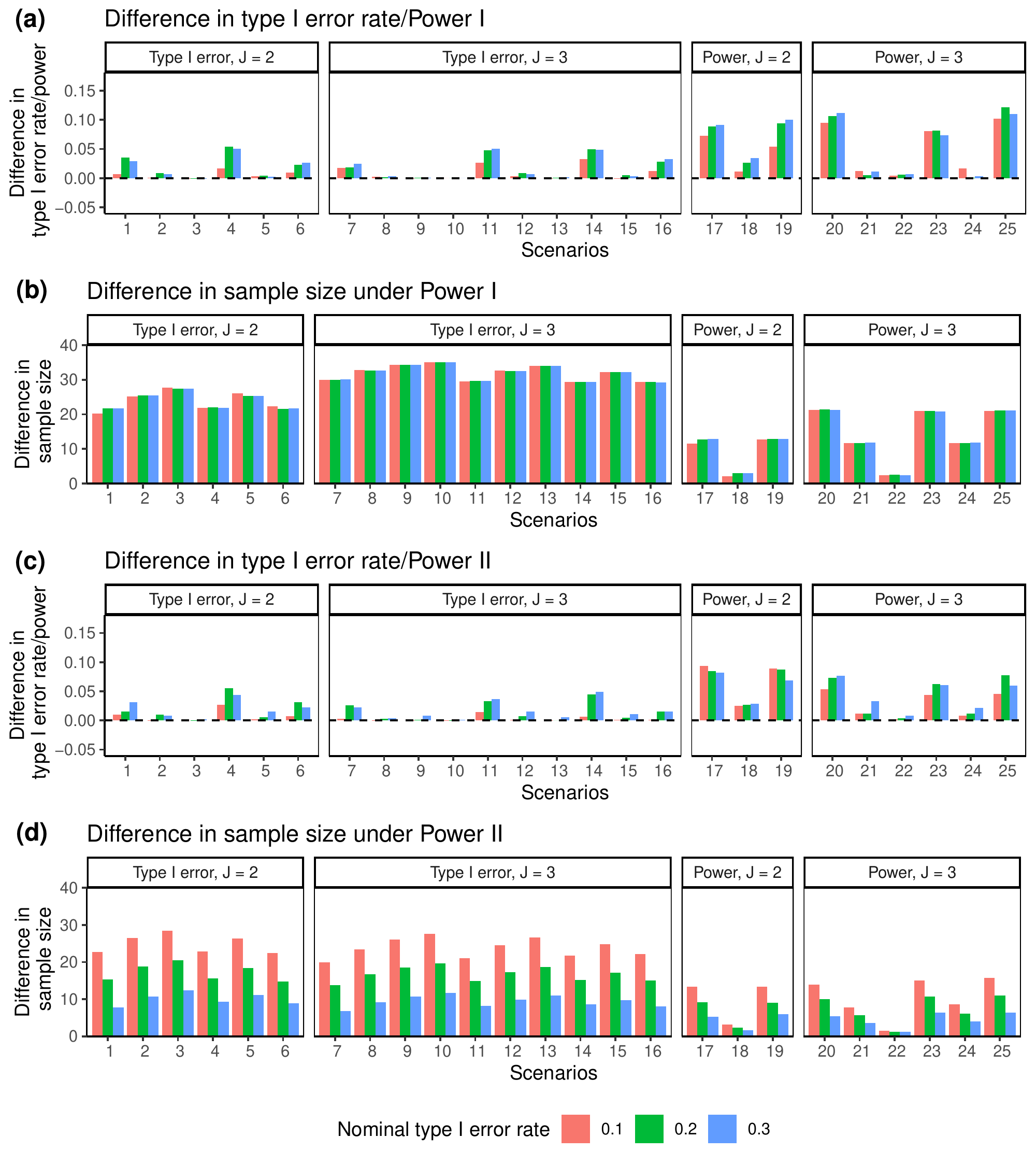}
    \caption{The difference in type I error/power and sample size between without and with two interim toxicity and futility monitoring when $(\phi_{T,0}, \phi_{T,1}) = (0.4, 0.2)$, $(\phi_{E,0}, \phi_{E,1}) = (0.2, 0.4)$, and $\alpha^*=0.1, 0.2$ or 0.3. Panels (a) and (b) show the results for $\beta_1^*=0.8$, and panels (c) and (d) show the results for $\beta_2^*=0.8$. }
    \label{fig:2IT_supp}
\end{figure}


\begin{figure}[H]
    \centering
    \includegraphics[width=0.9\textwidth]{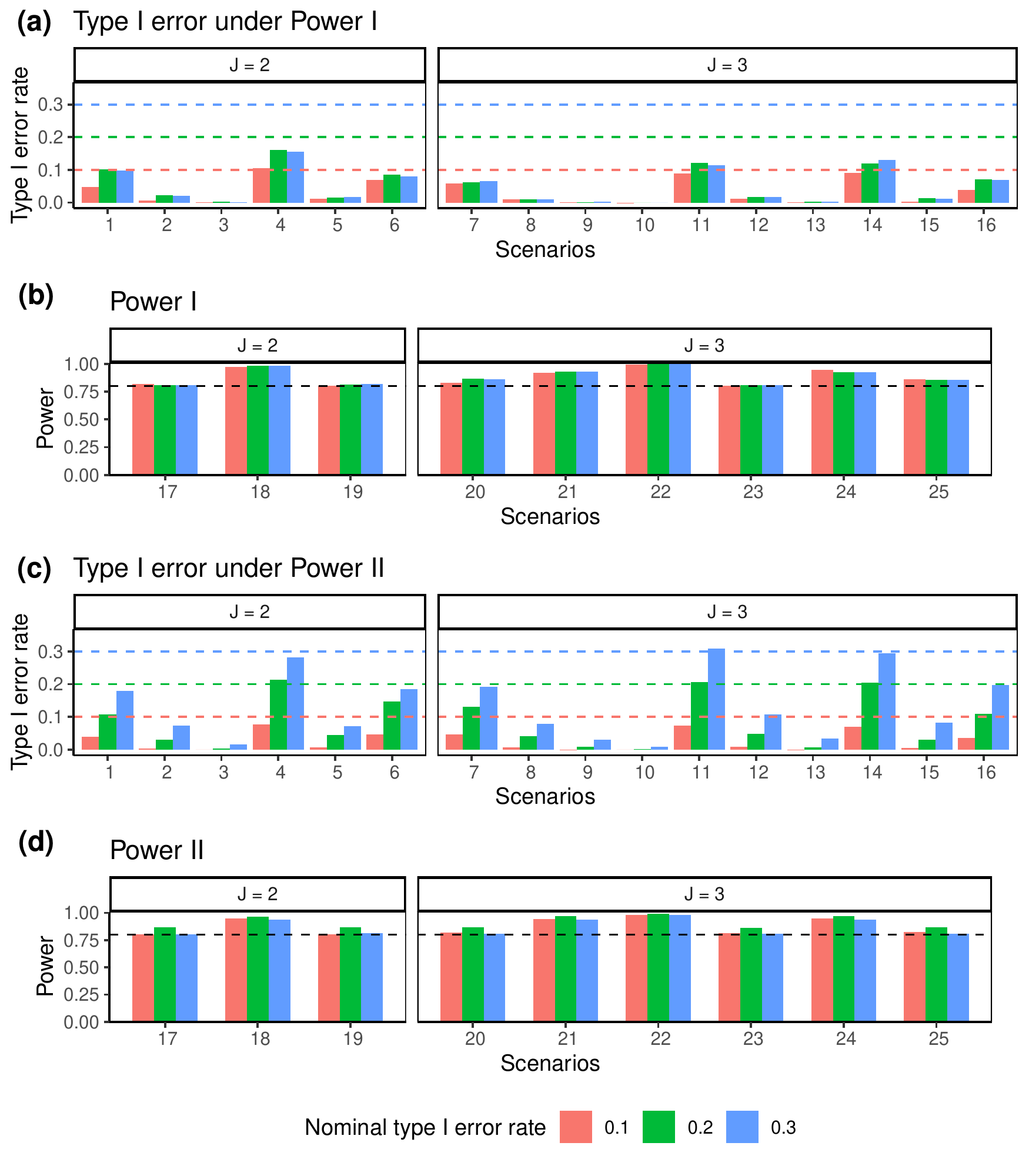}
    \caption{Results of sensitivity analysis when the correlation between $Y_E$ and $Y_E$ is $\rho=0.25$. Panels (a) and (b) show type I error and power  for $\beta_1^*=0.8$, and panels (c) and (d) show type I error and power for $\beta_2^*=0.8$. The other simulation parameters are $(\phi_{T,0}, \phi_{T,1}) = (0.4, 0.2)$, $(\phi_{E,0}, \phi_{E,1}) = (0.2, 0.4)$, and $\alpha^*=0.1, 0.2$ or 0.3. The horizontal dashed lines represent the nominal values of type I error and power. The toxicity and efficacy probabilities of scenarios 1-25 are given in Table 1. }
    \label{fig:OC_supp_rho_0.25}
\end{figure}


\begin{figure}[H]
    \centering
    \includegraphics[width=0.9\textwidth]{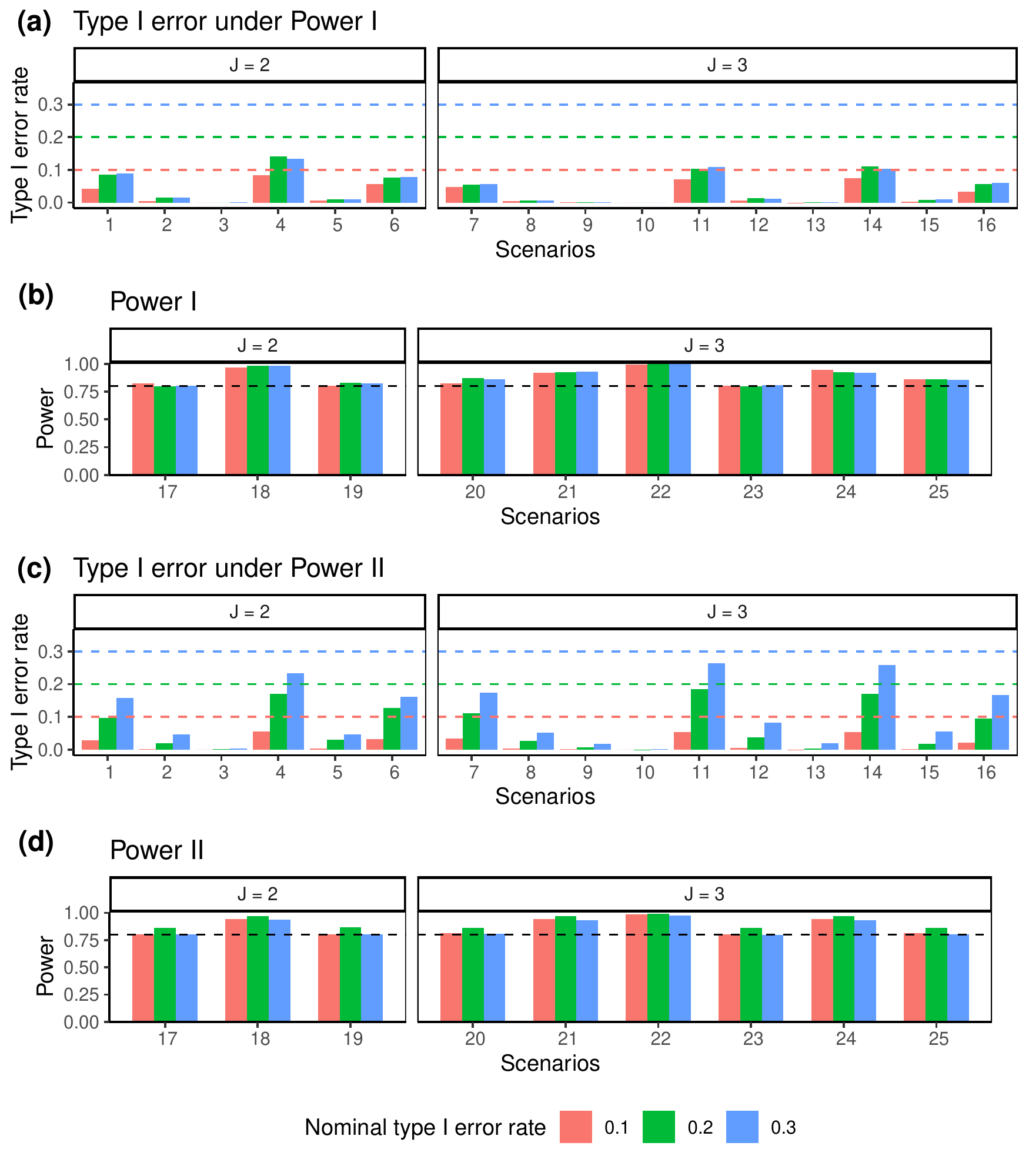}
    \caption{Results of sensitivity analysis when the correlation between $Y_E$ and $Y_E$ is $\rho=0.75$. Panels (a) and (b) show type I error and power for $\beta_1^*=0.8$, and panels (c) and (d) show type I error and power for $\beta_2^*=0.8$. The other simulation parameters are $(\phi_{T,0}, \phi_{T,1}) = (0.4, 0.2)$, $(\phi_{E,0}, \phi_{E,1}) = (0.2, 0.4)$, and $\alpha^*=0.1, 0.2$ or 0.3. The horizontal dashed lines represent the nominal values of type I error and power. The toxicity and efficacy probabilities of scenarios 1-25 are given in Table 1. }
    \label{fig:OC_supp_rho_0.75}
\end{figure}

\end{document}